\documentclass[useAMS,usenatbib]{mn2e}
\usepackage{amsmath}
\usepackage{graphicx}
\usepackage{subcaption}
\usepackage{float}
\usepackage{natbib}
\usepackage{etoolbox}

\title[The Star Formation Efficiency of Galaxies During the Epoch of Reionization]{Constraints on the Star Formation Efficiency of Galaxies During the Epoch of Reionization}
\author[G. Sun and S. R. Furlanetto]{G. Sun$^{1,2}$\thanks{E-mail: gsun@caltech.edu (GS); sfurlane@astro.ucla.edu (SF)} and S. R. Furlanetto$^{1}$\footnotemark[1] \\
$^{1}$Department of Physics and Astronomy, University of California, Los Angeles, CA 90024, USA \\
$^{2}$Cahill Center for Astronomy and Astrophysics, California Institute of Technology, Pasadena, CA 91125, USA}
\begin{document}

\date{\today}

\pagerange{\pageref{000}--\pageref{000}} \pubyear{0000}

\maketitle

\label{firstpage}

\begin{abstract}
Reionization is thought to have occurred in the redshift range of $6 < z < 9$, which is now being probed by both deep galaxy surveys and CMB observations. 
Using halo abundance matching over the redshift range $5<z<8$ and assuming smooth, continuous gas accretion, we develop a model for the star formation efficiency $f_{\star}$ of dark matter halos at $z>6$ that matches the measured galaxy luminosity functions at these redshifts. We find that $f_{\star}$ peaks at $\sim 30\%$ at halo masses $M \sim 10^{11}$--$10^{12}$~M$_\odot$, in qualitative agreement with its behavior at lower redshifts. We then investigate the cosmic star formation histories and the corresponding models of reionization for a range of extrapolations to small halo masses. We use a variety of observations to further constrain the characteristics of the galaxy populations, including the escape fraction of UV photons. Our approach provides an empirically-calibrated, physically-motivated model for the properties of star-forming galaxies sourcing the epoch of reionization. In the case where star formation in low-mass halos is maximally efficient, an average escape fraction $\sim0.1$ can reproduce the optical depth reported by Planck, whereas inefficient star formation in these halos requires either about twice as many UV photons to escape, or an escape fraction that increases towards higher redshifts. Our models also predict how future observations with JWST can improve our understanding of these galaxy populations. 
\end{abstract}

\begin{keywords}
early universe, reionization--galaxies: evolution--galaxies: high-redshift
\end{keywords}


\section{Introduction}

The epoch of reionization, occurring approximately 500 million to 1 billion years after the Big Bang ($6<z<15$), was the last major phase transition in the history of the universe, during which the neutral intergalactic medium (IGM) gradually transformed into the highly ionized state that we observe today. At the same time, the first stars and galaxies were forming from primordial gas clouds in the Universe. Although there is a clear connection between these two events (\citealt{BL_Review}; \citealt{BY_Review}; \citealt{L_F_2013BOOK}; \citealt{Robertson_HUDF}), their detailed relation is unknown, thanks to uncertainties in the properties of the galaxy populations. Fortunately, understanding the epoch of reionization itself will also shed light on the formation and evolution of early galaxies.

A number of diverse observational probes have helped us develop a preliminary picture of reionization over the past decade. Direct imaging of high-redshift star-forming galaxies has measured the abundance, luminosity distribution, and emission lines of galaxies out to $z\sim10$, from which valuable information about the budget of ionizing photons might be drawn (e.g., \citealt{Ellis_HUDF}; \citealt{Oesch_2015}; \citealt{Bouwens_LF15}; \citealt{Robertson_2015}; \citealt{Atek_2015}). 
Attempts have also been made to measure the ionization state of the IGM directly. Reionization is relatively well-constrained to be completed at $z\simeq6$ by analysis of the Gunn-Peterson effect \citep{G_P_1965} in the spectra of high-redshift QSOs (e.g., \citealt{Fan_2006}; \citealt{Willott_2007}; \citealt{Bolton_2011}; \citealt{McGreer_2015}) and gamma-ray bursts (GRBs, e.g., \citealt{Chornock_2013}). Preliminary measurements of the IGM neutral fraction at $z\sim7$--8 have been made using the Ly$\alpha$ emission fraction of star-forming galaxies (e.g., \citealt{Stark_2010}; \citealt{Treu_2013}; \citealt{Pentericci_2014}; \citealt{Tilvi_2014}; \citealt{Faisst_2014}; \citealt{Schenker_2014}). On the other hand, the integrated Thomson scattering optical depth, $\tau_{\rmn{es}}$, due to the scattering of CMB photons off free electrons after reionization, places an important integral constraint on the extended reionization history. The recently reported value $\tau=0.066\pm0.016$ by the \citet{Planck_Team} is significantly lower than $\tau=0.088\pm0.014$ previously measured by the \textit{Wilkinson Microwave Anisotropy Probe} (WMAP) satellite \citep{WMAP_Paper}.\footnote{The Planck measurement actually relies on a measurement of CMB lensing, as the $\tau$ estimate from the primary CMB anisotropies alone is somewhat higher \citep{Planck_July}.  Recently, \citet{Addison_2015} have questioned the self-consistency of the Planck parameter extraction. We will therefore comment on the implications of a higher optical depth as well.}
It is also necessary for these early ionizing sources to map smoothly onto their post-reionization counterparts at $z \la 5$, whose contribution to the global ionizing background has been characterized through measurements of the Ly$\alpha$ forest \citep{KF12,BB13}. 

Meanwhile, theorists have been trying to to develop plausible models for reionization that synthesize the results of these observations. For many years, the major challenge was reconciling the large optical depth observed by WMAP with the ionizing emissivity provided by galaxies at redshift $z>6$ \citep{Robertson_HUDF}. Both extrapolations of the estimated post-reionization ionizing background at $4 \la z \la 6$ \citep{B_H_2007, Calverley_2011} and measurements of the star formation rate evolution of Lyman break galaxies (LBGs, e.g., \citealt{Bouwens_LF12}; \citealt{Schenker_2013}; \citealt{Oesch_2015}) suggested a paucity of ionizing photons necessary for reproducing $\tau_{\rmn{es}}$. Resolving this dilemma required adjusting one or more of the many unknown parameters that map galaxy observations to reionization: the escape fraction of UV photons, $f_{\rm esc}$, the minimum halo mass to host a galaxy, $M_{\rm min}$, or other, more subtle parameters \citep{BY_Review, KF12}. For example, a common solution was to increase the overall ionizing emissivity by assuming an evolving escape fraction, $f_{\rmn{esc}}$, of ionizing photons (e.g., \citealt{Alvarez_2012}; \citealt{KF12}), which is motivated by both numerical simulations of star formation in the high-redshift universe \citep{W_C_2009, Ferrara_Loeb_2013} and the wide range of $f_{\rmn{esc}}$ values estimated from galaxy observations at lower redshifts \citep{Siana_2010, Nestor_2013, Mostardi_2015}. While the tension between the optical depth and ionizing emissivity has been largely resolved by Planck \citep{Robertson_2015, Bouwens_15Planck}, there remains a great deal of freedom in setting these important parameters of both galaxy formation and reionization. 

The primary goal of this paper is to investigate how current observations inform models of high-redshift galaxies and cosmic reionization. We will construct a simple analytic model of galaxy formation whose parameters can be constrained by existing observations but that also allows a theoretically-motivated extrapolation to earlier epochs. The halo abundance matching approach \citep{VO_2004}, which empirically associates galaxies and dark matter halos by matching their number densities, has been widely used to model the star formation activity in galaxies across cosmic time \citep{C_W_2009, Yang_2012, B_S_2015}. Taking advantage of its simplicity for analytic models, several recent studies have demonstrated its utility for empirically predicting the luminosity functions of high-redshift galaxies responsible for reionization \citep{Mashian_2016, Mason_2015, Visbal_2015}. 

In this paper, we will first give a more thorough analysis of the star formation efficiency implied by abundance matching, assuming continuous star formation in high-redshift galaxies. In contrast to most treatments of reionization, our approach allows the star formation efficiency $f_{\star}$ to evolve with both halo mass and redshift and thus effectively alters the overall ionizing efficiency. Whereas many analytic models simply treat $f_{\star}$ as an arbitrary constant, our approach carefully constrains $f_{\star}$ with abundance matching, providing a more accurate way of calibrating models of reionization to observations. Because it is calibrated to bright galaxies at $z \sim 6$--8, our approach will provide a ``baseline" model for as-yet-undiscovered galaxy populations during the ``Cosmic Dawn": while the extrapolation (to both higher redshifts and fainter luminosities) is by no means a firm theoretical prediction, it at least provides a clear understanding of the role of known galaxy populations in inferences about the reionization process. Here we explore these implications in some detail through comparisons of our extrapolations to several other observables, and we comment on how additional populations or effects (such as feedback or Population~III stars) may affect the results. 

This paper is organized as follows. In Section 2, we describe our simple analytic model for continuous mode star formation in high-redshift galaxies based on the dust-corrected, rest-frame UV luminosity functions and the mass assembly history of dark matter halos. We also provide a brief discussion of the halo abundance matching (HAM) technique. We compare our modeled star formation histories to observations of high-redshift galaxies and present our predictions for JWST in Section 3. Then, in Section 4, we show how to use our star formation models to calculate simple reionization histories. Section 5 is dedicated to a joint analysis of galaxy populations and observational constraints, in particular the optical depth measured by Planck. In Section 6, we discuss a few possible variations around our baseline model, including physical processes such as photo-suppression of low-mass galaxies and the formation of very massive Pop~III stars. Finally, we briefly conclude in Section 7. 

Throughout this paper, we assume a flat, $\rmn{\Lambda CDM}$ cosmology with $\Omega_{m}=0.28$, $\Omega_{b}=0.046$, $\Omega_{\Lambda}=0.72$, $\sigma_{8}=0.82$, $n_{s}=0.95$, and $h=0.7$, consistent with the most recent measurement from Planck \citep{Planck_Team}. We also assume a Salpeter initial mass function (IMF) from 0.1 to 100 $M_{\odot}$ and a stellar metallicity of $Z_{*}=0.05Z_{\odot}$ for calculations of star formation and reionization. All the magnitudes cited are expressed in terms of $m_{\rmn{AB}}=-2.5 \log_{10} (f_{\nu}/\rmn{nJy}) + 31.40$ \citep{Oke_1974} and the base-10 logarithm is assumed unless stated otherwise.


\section[]{Modeling the Cosmic Star Formation History}

We use a simple but physically intuitive approach to model the average star formation rate (SFR) inside a galaxy at redshift $z\ga5$. When the timescale of star formation is much less than the dynamical time on which galaxies grow (i.e. stars form ``instantaneously"), the star formation rate $\dot{M}_{\rmn{SFR}}$ can be approximated as a balance between the gas fueling rate, $\dot{M}_{\rmn{acc}}$, and the rate, $\dot{M}_{\rmn{W}}$, at which galactic outflows deplete the gas \citep{Munoz_2012}. We introduce a free parameter, $f_{\star}$, which measures the star formation efficiency, namely, the fraction of accreted baryons that forms stars:\footnote{Note that $f_\star$ is defined in an ``instantaneous" sense, proportional to the current accretion rate and depending on the current halo mass and redshift. As a halo grows, $f_\star$ will therefore change, and our results require averaging over halo growth histories in order to calculate a halo's \emph{net} star formation efficiency.} 
\begin{equation}
	\dot{M}_{\rmn{SFR}} = \dot{M}_{\rmn{acc}} - \dot{M}_{\rmn{W}} = f_{\star} \dot{M}_{\rmn{acc}}  = f_{\star} \frac{\Omega_{\rmn{b}}}{\Omega_{\rmn{m}}} \dot{M}_{h}, 
	\label{eq:def_f}
\end{equation}
where $\dot{M}_{h}$ is the total halo accretion rate. 

As a result, the star formation history of galaxies can be directly related to the mass assembly history of dark matter halos. The evolution of $\dot{M}_{h}$ has been studied using catalogs of halos from cosmological simulations (see Section \ref{acc-hist}), allowing us to describe the SFR once $f_{\star}$ can be characterized. For high-redshift galaxies dominated by massive, young stars, the rest-frame UV 1500~$\rmn{\AA}$ luminosity $L_{\rmn{UV,1500}}$ is a good indicator of the SFR if any significant fluctuations in the star formation rate occur on a timescale longer than the evolution timescale of the massive stars, which is about 100 Myr \citep{Kennicutt_1998, M_D_2014}, and provided that dust attenuation is properly accounted for. Specifically, the dust-corrected SFR of a galaxy is proportional to its rest-frame UV continuum ($1500 \, \rmn{\AA}$--$2800\, \rmn{\AA}$) luminosity by 
\begin{equation}
	\dot{M}_{\rmn{SFR}} = \mathcal{K}_{\rmn{UV,1500}} \times L_{\rmn{UV,1500}},
\end{equation} 
where we assume a fiducial constant $\mathcal{K}_{\rmn{UV,1500}} = 1.15 \times 10^{-28} ~\rmn{M_{\odot}}~\rmn{yr}^{-1}/~\rmn{ergs}~\rmn{s}^{-1}~\rmn{Hz}^{-1}$ evaluated for continuous mode star formation with a Salpeter IMF.\footnote{The adopted $\mathcal{K}_{\rmn{UV,1500}}$ is a compromise value assuming an evolving $Z_{*}=10^{-0.15z}Z_{\odot}$, $\sim20\%$ smaller than the value from \cite{Kennicutt_1998} which assumes solar metallicity (see \citealt{M_D_2014} for details).} The factor $\mathcal{K}_{\rmn{UV,1500}}$ introduces an overall uncertainty in the normalization of $f_\star$, due to its dependence on the unknown IMF and metallicity of these stars (or a systematic difference if the conversion is mass or redshift dependent). Note that we make the necessary conversions when comparing to other models. 

\subsection{Properties of Dark Matter Halos}  \label{acc-hist}

\begin{figure}
 \includegraphics[width=0.45\textwidth]{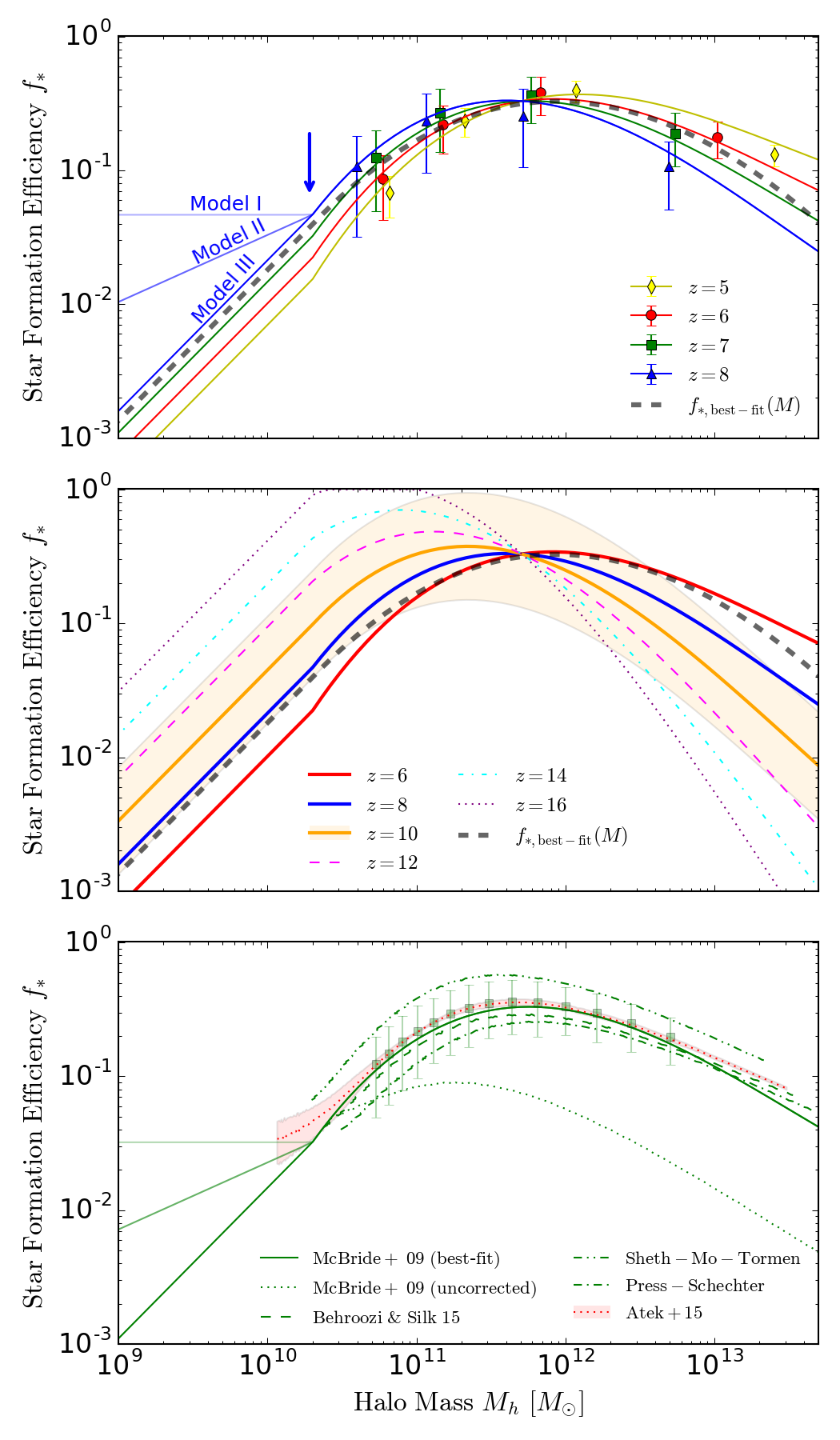}
\caption{ Star formation efficiency $f_{\star}$ as a function of halo mass and redshift. \emph{Top}: Fits in the observed redshift range ($z=5$--8). Data points with error bars show the $1\sigma$ uncertainties from 1000 Monte Carlo fits to the observed Schechter functions. Solid curves show the $z$-dependent fits to the data, while the thick dashed curve shows the $z$-independent fit to the same data. The vertical arrow indicates the detection limit of the UDF12 survey at $z=8$ ($M_{\rmn{lim}}\sim-18$; this increases slightly toward lower redshifts). Below this limit, we show three potential extrapolations of the abundance matching relation (see text). \emph{Middle:} Solid cures show the evolution of $f_\star(M,z)$ in our $z$-dependent fit, assuming the extrapolation of Model III. \emph{Bottom:} Variations in model results, shown at $z=7$. The curves vary the input assumptions (see legend and text). The red shaded region shows $f_\star$ inferred from \citet{Atek_2015} and $M_{\rmn{lim}}=-15.25$.}
 \label{fig:f_star_sc}
\end{figure}

The halo mass function $n(M)$ measured from cosmological N-body simulations is usually expressed in the form
\begin{equation}
	n(M) = \frac{\mathrm d N}{\mathrm d M} = \frac{\bar{\rho}_{m}}{M} f(\sigma) \left|\frac{\mathrm d \log \sigma}{\mathrm d M}\right|,
\end{equation}
where $\bar{\rho}_{m}$ is the average matter density and $\sigma$ is the density variance smoothed over the scale $M$. $f(\sigma)$ is a function determined by the particular fit (or often through analytic arguments, as in the original \citealt{P_S_1974}). Here we adopt the \citet{Tinker_2008} mass function, corrected for the high-redshift universe by \cite{Behroozi_2013} and implemented by the online calculator HMF\textit{calc}\footnote{http://hmf.icrar.org} \citep{Murray_2013}. We note that using different functional forms does change our parameterization of the star formation efficiency $f_{\star}$ slightly (see Fig.~\ref{fig:f_star_sc}), but given the close agreement with high-redshift observations shown in \cite{Behroozi_2013} and the much greater uncertainties arising from the luminosity function, we do not regard the mass function as a systematic worry in our final results. 

Many empirical and analytical models have been developed to characterize the growth history of dark matter halos \citep{McBride_2009, B_S_2015, Correa_2015}. Noting the general similarity among their predicted mass accretion histories and accretion rates, we adopt the two-parameter model given by \cite{McBride_2009}, which fits the halo's baryonic mass accretion rate as
\begin{equation}
	\dot{M}_{\rmn{acc}} \approx 3\ \rmn{M_{\odot}/yr} \left(\frac{M_{h}}{10^{10} M_{\odot}}\right)^{1.127} \left(\frac{1+z}{7}\right)^{\eta},
	\label{eq:acc-rate}
\end{equation} 
where \citet{McBride_2009} and \citet{Goerdt_2015} found $\eta = 2.5$ at high redshifts. The $z$-dependence can be understood from the redshift scaling of the dark matter halo mass function \citep{Dekel_2013}. The mass dependence has only been tested in simulations over a limited range of halo masses (typical of galaxies), and the power-law assumption is probably not a good assumption for very small or very large halos. However, over the limited range of halo masses relevant to galaxy formation at $z \sim 6$ (where the mass function is very steep), we have found a power-law approximation to be reasonably accurate.

We note that our continuous accretion model averages over galaxy mergers, which can affect the inferred star formation efficiency \citep{Tissera_2000, D_A_2008, C_W_2009, Pawlik_2011, Vulcani_2016}. \citet{Goerdt_2015} have shown that, over a broad range of halo masses, a clear majority of accretion occurs through the continuous mode, at least at moderate redshifts, while \citet{B_S_2015} have found similar results at higher redshifts. According to equation~(\ref{eq:acc-rate}), the mean mass accretion rate increases much more rapidly than the Hubble expansion rate toward high redshifts, so halos grow very rapidly which likely decreases the stochasticity due to mergers. Below we will allow for some scatter in the halo mass-luminosity relationship, which can partly be due to the effects of mergers. Fortunately, its presence does not affect our qualitative results (and has only a small quantitative effect), though a model relying entirely on mergers shifts the overall abundance matching relation substantially \citep{Visbal_2015}. Nevertheless, because the merger rate increases slowly with halo mass \citep{B_S_2015}, mergers may affect our inferences about the brightest galaxies, and we intend to examine their effects in more detail in future work. 

To determine the minimum halo mass in which galaxies can form, $M_{\rm min}$, we take the criterion given by \citet{Okamoto_2008}, which takes account of galaxy mass loss by incorporating a spatially uniform and time dependent UV background in their cosmological hydrodynamical simulations. In their model, $M_{\rmn{min}}$ is evaluated using the equilibrium temperature $T_{\rmn{eq}}$ of the gas at the edge of the halo as \citep{N_M_2014}
\begin{equation}
	M_{\rmn{min}}(z) = \frac{1}{G H_{0}} \left(\frac{2k_{B}T_{\rmn{eq}}(\Delta_{\rmn{vir}}/3)}{\mu m_{p}(1+z)}\right)^{3/2} \left(\frac{2 \Omega_{m}(z)}{\Delta_{c}(z)\Omega_{m,0}}\right)^{1/2}, 
	\label{eq:M_min}
\end{equation} 
where $\Delta_{\rmn{vir}}$ is the halo's virial density, $\Delta_{c}(z)=18\pi^{2}+82d-39d^{2}$ and $d=\Omega_{m}(z)-1$. 

\subsection{Halo abundance matching}

Given that the UV luminosity traces the star formation rate of galaxies (and hence accretion onto dark matter halos, according to our model), we wish to map the UV luminosity function onto the mass function of dark matter halos. We use the halo abundance matching (HAM) technique \citep{VO_2004} to assign a unique halo mass $M_{h}$ to each UV luminosity $L$ by solving the equation
\begin{equation}
	\int_L^{\infty} \phi(L) \mathrm d L = \int_{M_{h}(L)}^{\infty} n(M_{h}) \mathrm d M_{h},
\end{equation}
for $M_{h}(L)$, where $\phi(L)$ and $n(M_{h})$ are the luminosity function of galaxies 
(or more precisely the intrinsic luminosity function derived from the observed one after taking the scatter in the $L$--$M_{h}$ relation into account) and the mass function of dark matter halos respectively.\footnote{$\phi(L)$ is more commonly expressed in the magnitude form $\phi(M) = 0.4\ln10\phi_{*}\left[10^{0.4\left(M_{*}-M\right)}\right]^{1+\alpha} \exp \left[-10^{0.4\left(M_{*}-M\right)}\right]$, where $\phi_{*}$, $M_{*}$, and $\alpha$ are the normalization factor, the characteristic magnitude and the faint-end slope, respectively.} We take the latest Schechter parameterizations (see Table~\ref{tb:Schechter_Params}) of the observed UV luminosity functions at redshift $z=5$ to $z=8$ from \citet{Bouwens_LF15}; \citet{Atek_2015} have recently gone deeper at $z \sim 7$ using strongly lensed clusters, and we will comment on the implications of their results as well. It is also worth noting that \citet{Bowler_2015} find that a Schechter function might underestimate the bright end of the luminosity function at $z\sim6$ and thus might not be the best functional form at high redshifts. However, other studies find it an acceptable fit, and the bright end of the luminosity function does not dominate the integral quantities most relevant to reionization, so we simply adopt the Schechter form in this paper. Throughout these calculations, we use the observed detection limits from the UDF12 program ($M_{\rm lim} \approx -17.7$) to determine the minimum luminosity in our abundance matching. 

Research suggests that dust attenuation is non-trivial in galaxies at $5<z<8$ \citep{Smit_2012, Bouwens_2014, Finkelstein_2015}, so we apply the method introduced by \citet{Smit_2012} to perform dust corrections on the observed luminosity functions using the well-established relation between the UV continuum slope $\beta$ and $M_{\rmn{UV}}$ \citep{Bouwens_2014} and the linear fit of infrared excess $A_{1600}$ versus $\beta$ established by \cite{Meurer_1999}. We note, however, that the latter relation was calibrated by starburst galaxies in local universe and therefore its application to galaxies at redshift $z\ga6$ is uncertain. 

\begin{table}
 \caption{Schechter parameters derived for the rest-frame UV luminosity functions at redshift $z\sim5$, $z\sim6$, $z\sim7$ and $z\sim8$}
 \label{tb:Schechter_Params}
 \begin{tabular}{@{}cccc}
 \hline
 \hline
 $\langle z \rangle$ & ${M_{\rmn{UV}}^{*}}^{\rmn{i}}$ & $\phi^{*}$ ($10^{-3}$\ $\rmn{Mpc^{-3}}$) & $\alpha$ \\
 \hline
 4.9 & $-21.17 \pm 0.12$ & $0.74^{+0.18}_{-0.14}$ & $-1.76 \pm 0.05$ \\
 5.9 & $-20.94 \pm 0.20$ & $0.50^{+0.22}_{-0.16}$ & $-1.87 \pm 0.10$ \\
 6.8 & $-20.87 \pm 0.26$ & $0.29^{+0.21}_{-0.12}$ & $-2.06 \pm 0.13$ \\
 7.9 & $-20.63 \pm 0.36$ & $0.21^{+0.23}_{-0.11}$ & $-2.02 \pm 0.23$ \\
 \hline
 \end{tabular}

 \medskip
$^{\rmn{i}}$ These are determined at a rest-frame wavelength 1600$\rmn{\AA}$. 
\end{table}

\begin{figure}
 \includegraphics[width=0.5\textwidth]{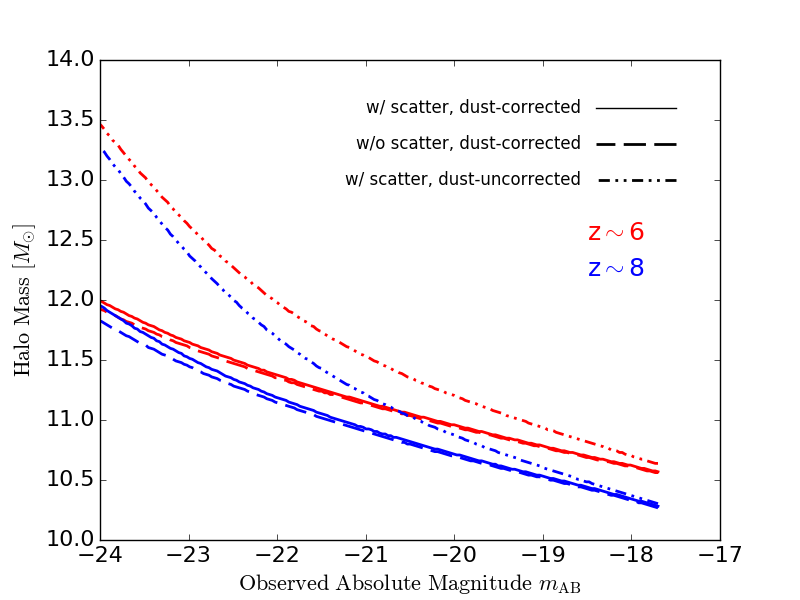}
 \caption{Effects of the dust correction and stochasticity in the $L$--$M_{h}$ relation on the derived mass-magnitude relation from halo abundance matching. The solid curves show our full model, while the dashed and dot-dashed curves ignore the scatter and dust correction, respectively. 
A limiting magnitude of -17.7 mag is assumed in light of current detection limit.}
 \label{fig:ham}
\end{figure}

The traditional abundance matching technique assumes that the galaxy luminosity or stellar mass is a monotonic function of halo mass, and vice versa. 
This ignores the possibility of intrinsic scatter in the galaxy-dark matter halo relation as well as complications from satellite galaxies  \citep{Behroozi_2010, Moster_2010}. The latter appears to be a small effect \citep{Wang_2006, Moster_2010}, and at high redshifts it is difficult to observationally distinguish satellite and central galaxies thanks to the very small sizes of the halos. To incorporate the former effect, we adopt the deconvolution method described in \citet{Behroozi_2010}, which assumes the distribution $P_{s}(L|M_{\rmn{h}})$ to be a lognormal in $L$
\begin{equation}
	P(L|M_{\rmn{h}}) \mathrm d L = \frac{1}{\sqrt{2\pi} \sigma_{\log L} \ln10} \exp \left[ -\frac{\log L - \langle \log L \rangle}{2\sigma_{\log L}^{2}}\right] \frac{\mathrm d L}{L}, 
\end{equation}
with $\sigma_{\rmn{\log L}}$ being independent of the halo mass. The observed luminosity function $\phi_{\rmn{obs}}$ described by the Schechter parameters can therefore be related to an unknown intrinsic luminosity function $\phi_{\rmn{intrinstic}}$ by solving the convolution
\begin{equation}
	\phi(L)_{\rmn{obs}} = \int_{-\infty}^{\infty} \phi(10^{x})_{\rmn{intrinstic}} P_{s}(x - \log L) \mathrm d x. 
\end{equation}
We note that the logarithmic spread in galaxy luminosity at a given halo mass, $\sigma_{\log L}$, is determined to be a constant $\sim0.2$ both analytically \citep{McBride_2009, Munoz_2012} and observationally \citep{More_2009}, so we take $\sigma_{\log L}=0.20\pm0.05$. 

Fig.~\ref{fig:ham} shows the halo mass versus the galaxy's observed absolute magnitude given by the abundance matching approach, as well as how the relationship depends on the dust correction and $\sigma_{\log L}$. Here the solid curves show our full model. The dash-dotted curves assume no dust correction. In particular, the dust correction is larger for larger halo masses (i.e. brighter galaxies), resulting in a net flattening of the observed luminosity function. Abundance matching without the luminosity scatter $\sigma_{\log L}$ is represented by the dashed curves. As the scatter flattens the bright end of the intrinsic luminosity function, abundance matching to the observed luminosity function directly slightly underestimates the $M_{h}$ at a given luminosity.

\subsection{Evolution of $f_{\star}(M,z)$}

Fig.~\ref{fig:f_star_sc} shows the star formation efficiency, $f_{\star}$, as a function of halo mass and redshift, that results from combining equations~(\ref{eq:def_f}) to~(\ref{eq:M_min}). Here the uncertainties in $f_{\star}$ are determined from the quoted errors on the three Schechter parameters and $\sigma_{\log L}$ via 1000 Monte Carlo realizations. We note that the error bars provided represent the uncertainty at a specific halo mass estimated from Monte Carlo realizations and do not consider the correlation between data points resulting from abundance matching. We also ignore correlations between the measured Schechter parameters for simplicity, finding that they can change the inferred shape marginally but remain within the uncertainty envelope illustrated here. The impact of choosing different mass functions and halo accretion models is insignificant compared with the observationally-driven uncertainties in $f_{\star}$. However, including the dust correction raises $f_{\star}$ by a factor of $\sim4$ and slightly increases the peak halo mass, since the former is related to the amplitude of $L/M_{h}$ whereas the latter is determined by the slope $\mathrm d \log L/ \mathrm d \log M_{h}$ (see Fig.~\ref{fig:ham}). 

Fig.~\ref{fig:f_star_sc} shows that the star formation efficiency evolves strongly with halo mass, peaking at a characteristic halo mass of $\sim10^{11}$--$10^{12}M_{\odot}$ and decreasing at both smaller and larger masses. At lower redshifts, similar relationships have also been identified (\citealt{C_W_2009}; \citealt{Moster_2010}; \citealt{Kravtsov_2014}; \citealt{B_S_2015}). At these times, AGN feedback and feedback from massive stars, such as supernovae and stellar winds, are considered to be the dominant physical processes that suppress the star formation at the high- and low-mass ends, respectively (\citealt{Mo_book}; \citealt{Kravtsov_2014}). At high redshifts, these feedback mechanisms are likely to be less efficient, because (at a fixed halo mass) galaxies are more compact at those times. However, their effects on $z>5$ galaxies are highly uncertain.

Additionally, there might be a non-trivial dependence on redshift, although it is difficult to quantify the dependence given the large errors, especially for smaller masses. The halo mass at which $f_*$ peaks evolves with redshift as 
$\log_{10}(M_{\rmn{peak}}/M_{\odot})=-0.15(1+z)+12.8 $ for $5<z<10$, which is broadly consistent with that found by \cite{B_S_2015}. This empirically-derived trend might indicate evolution in the feedback mechanisms that control $f_\star$. However, there is no obvious reason for $M_{\rm peak}$ to decrease toward higher redshifts, as the two most important factors in decreasing the efficiency in massive halos at low redshifts (AGNs and heating at the virial shock) are probably less important at high redshifts.


\begin{figure}
 \includegraphics[width=0.5\textwidth]{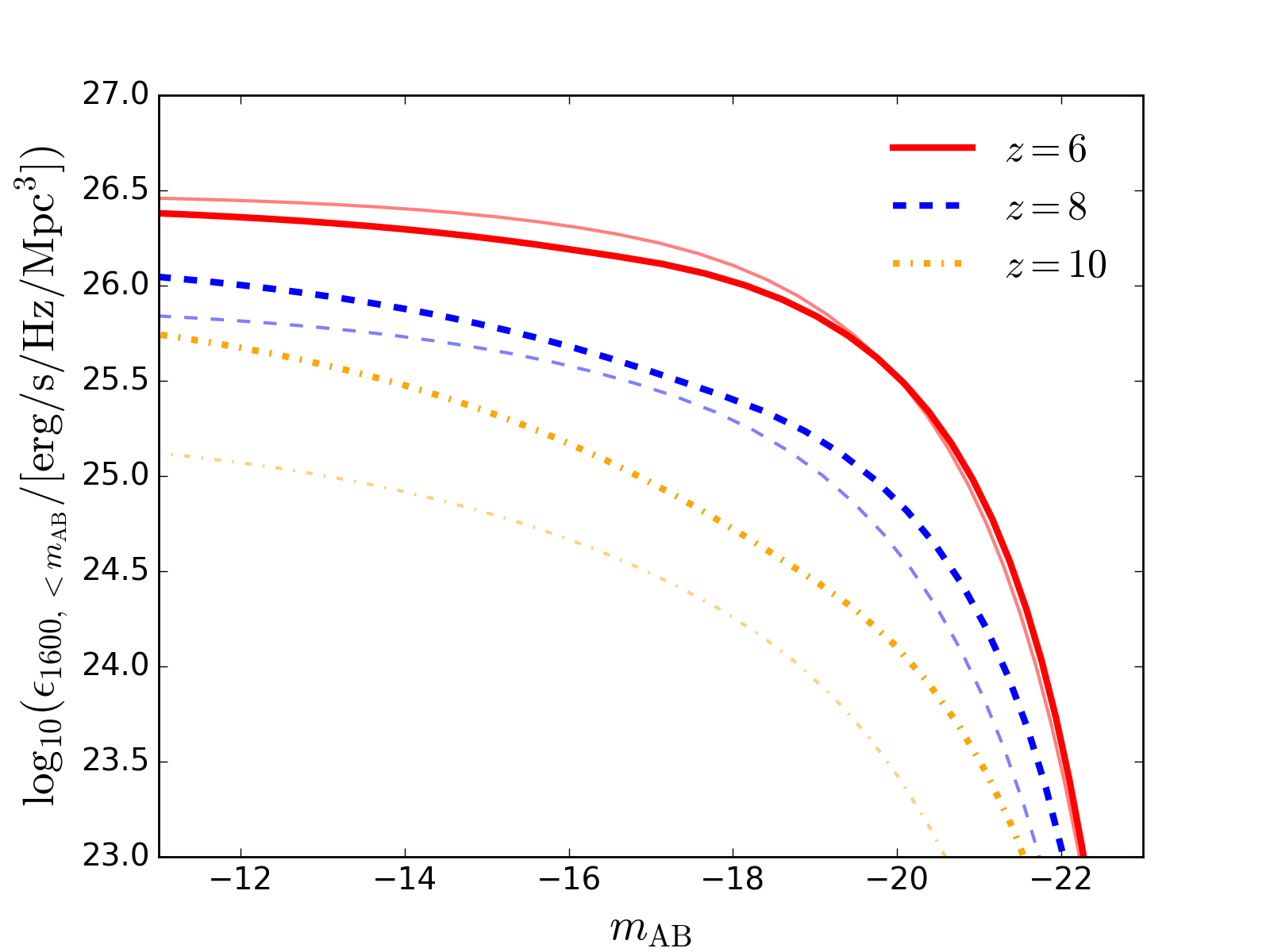}
 \caption{Cumulative emissivity as a function of UV magnitude. The thick, opaque curves show the emissivities integrated down to the corresponding magnitude ($x$-axis), given a minimum halo mass of $\sim10^{8}M_{\odot}$ and our fiducial $z$-dependent model (Model~II, see later text). The thin, transparent curves represent the same emissivities but assuming $f_{\star}$ does not evolve with redshift. }
 \label{fig:Emissivity}
\end{figure}

More generally, we can examine the overall dependence of $f_\star$ on both mass and redshift. Fig.~\ref{fig:f_star_sc} shows our fiducial, six-parameter multiple regression fit 
\begin{align}
	\rmn{log}f_{\star} =\ & a_{0} + a_{1} \rmn{log}M + a_{2} (1+z)/8 + a_{3} [(1+z)/8] \rmn{log}M \nonumber \\
	& + a_{4} (\rmn{log}M)^{2} + a_{5} (\rmn{log}M)^{3}, 
	\label{eq:fit_f_mz}
\end{align}
to the abundance matching results, illustrated at 
$z=5$--8 in the top panel. The best-fit parameters are $a_{0}=-178.17\pm24.40, a_{1}=39.11\pm6.10, a_{2}=10.74\pm0.79, a_{3}=-0.92\pm0.07, a_{4}=-2.87\pm0.51$ and $a_{5}=0.07\pm0.01$. We intentionally avoid higher-order terms in $z$ in order to prevent strong redshift evolution from appearing as an artifact of our fits. 
Fig.~\ref{fig:f_star_sc} shows the evolution in this model from $z = 6$--16. All the curves intersect near $10^{11.5}M_{\odot}$, where the two redshift-dependent terms cancel with each other, leaving the expression independent of redshift. This is nothing but an artifact of the functional form, which models the redshift evolution to the linear order. 

For comparison, the thick dashed curve in the top and middle panels of Fig.~\ref{fig:f_star_sc} shows the best-fit result from a regression of the same order in $\log M$, assuming no redshift dependence (i.e. setting $a_{2}$, $a_{3}$ to 0). The key effects of our fitted form's redshift dependence are to decrease $M_{\rm peak}$ and to steepen the decline toward higher masses as $z$ increases. The bright end of the luminosity function is therefore relatively uncertain at high redshift, but \emph{integrated} quantities like the global emissivity, which depend primarily on faint galaxies, do not evolve as rapidly.  As shown in Fig.~\ref{fig:Emissivity}, the cumulative emissivity is dominated by galaxies well below the characteristic magnitude ($\sim-21$) at all redshifts, whether or not we allow redshift evolution. However, the $z$-dependent model predicts a much gentler decline in the total emissivity toward higher redshifts, thanks to the decrease of $M_{\rm peak}$ with $z$. For completeness, we also evaluate the importance of the assumed linear redshift dependence (at a fixed $M_{h}$) in equation~(\ref{eq:fit_f_mz}). We find that the uncertainties at $z\sim15$ induced by deviations from our assumed dependence are comparable in amplitude to the envelope on the $z=10$ curve shown in Fig.~\ref{fig:f_star_sc}. 

\begin{figure*}
 \includegraphics[width=0.8\textwidth]{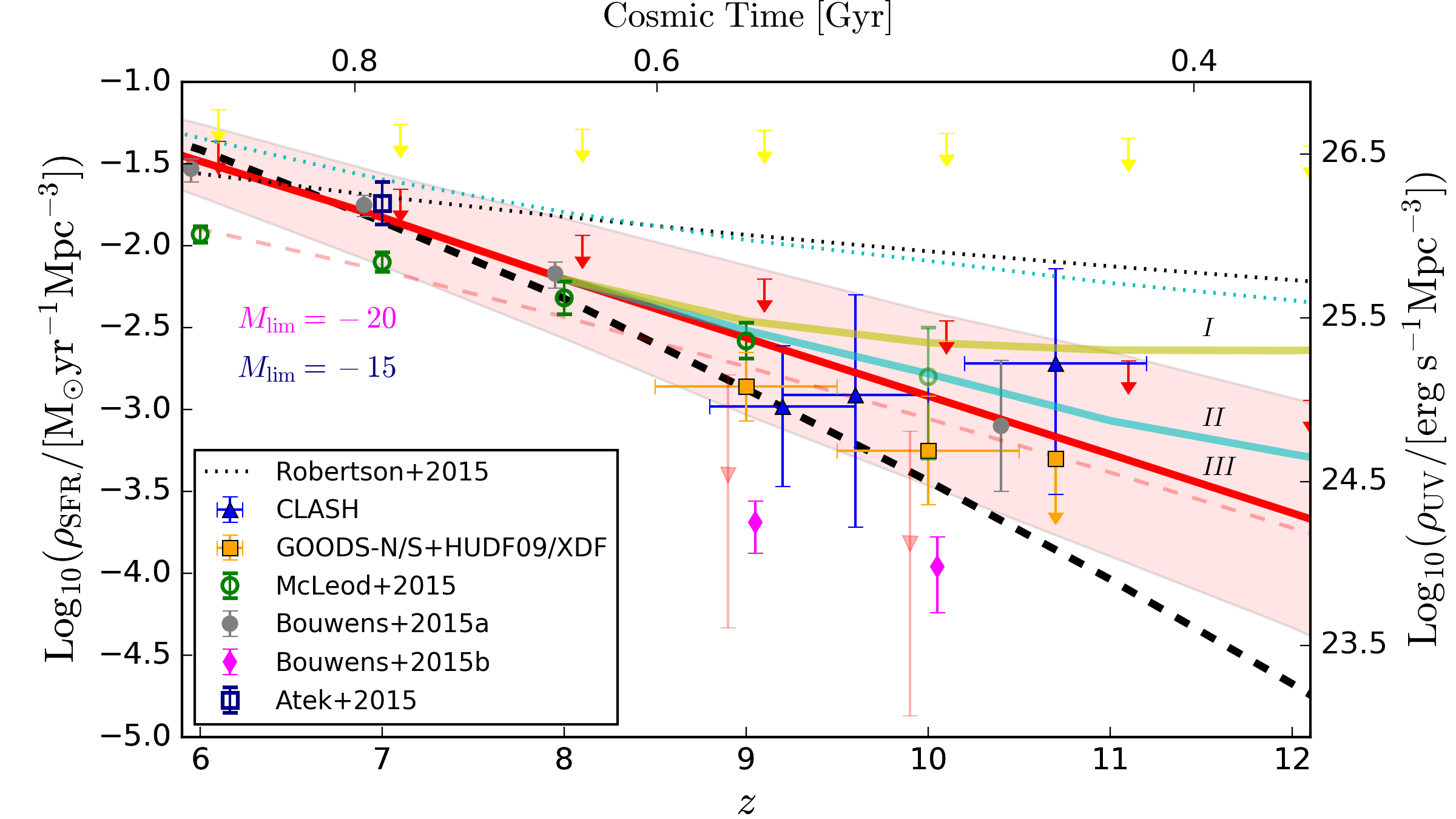}
 \caption{The evolution of the star formation rate density (SFRD) and the UV luminosity density with redshift.
The red solid curve shows the SFRD evolution calculated by Model~III, with the shaded region representing the 68\% confidence interval, integrated down to $M_{\rmn{lim}}=-17.7$ to match observations from Hubble Frontier Field data, shown by the green (dust-uncorrected, \citealt{McLeod_2015}) and grey (dust-corrected, \citealt{Bouwens_LF15}a) circles respectively. Cases where the dust correction is not included (thin dashed curve), Model I/II is assumed (yellow/cyan curves, only the deviated parts from Model II), or $z$-evolution is ignored (thick dashed curve) are shown for comparison. As mentioned in the text, the mass cutoff $2 \times 10^{10}M_{\odot}$ distinguishing the three models is chosen so that all of them are consistent with the 1--$\sigma$ interval of the observed $\rho_{\rmn{SFR}}$ at $z \la 8$. The pink diamonds represent the SFRD derived from the observed bright end of UV luminosity functions at $z\sim9$ and 10 (\citealt{Bouwens_2015_z910}b), whereas the red inverted triangles show our predictions. The black, dotted line shows the maximum-likelihood fit to $\rho_{\rmn{SFR}}$ given by \protect\cite{Robertson_2015}, assuming a limiting luminosity $L>0.001L_{*}$ (or $M_{\rmn{lim}}\sim-13$), whereas the cyan, dotted line represents the prediction by Model II at the same limiting magnitude. The upper limits shown illustrate the evolution of the maximum possible SFRD calculated by extrapolating our best-fit Model~I (yellow) and Model~III (red) down to the minimum halo mass $M_{\rmn{min}}$ given by \protect\cite{Okamoto_2008}.}
 \label{fig:SFRD}
\end{figure*}

As star formation might be sustained in halos of mass as low as $\sim10^{8}M_{\odot}$, the cumulative star formation history is sensitive to the extrapolation of $f_{\star}$ with mass as well. In contrast with similar works which treat $f_{\star}$ as a redshift-independent parameter (e.g. \citealt{Mashian_2016}; \citealt{Mason_2015}), we fit $f_{\star}$ as a function of both halo mass and redshift. We then explore three possible scenarios for extrapolating toward the dominant low masses: (i) \emph{Model I} assumes a constant $f_{\star}$ below the mass cutoff $M=2 \times 10^{10}M_{\odot}$;\footnote{The value $2 \times 10^{10}M_{\odot}$ is chosen here so that our extrapolations yield reasonable star formation rate densities at $5 \la z \la 10$ in conformity with observations, see Section 3 and Fig.~\ref{fig:SFRD} for details.} (ii) \emph{Model II} assumes a power-law extrapolation of slope 0.5 below the mass cutoff, and (iii) \emph{Model III} fits a power law to the low-mass end according to the average (i.e. $z$-independent) slope $\partial \log f_{\star}/\partial \log M \approx 1.1$ over $5 \la z \la 8$.\footnote{This mass dependence is considerably steeper than expected from the simplest theoretical models of supernova wind feedback (e.g., \citealt{Dave_2011, Dayal_2014}). We emphasize again that the slope is not well-constrained by observations.} Note that in Model~III, we use the average slope because the evolution of the cutoff mass and the low-mass end slope will bias the result of $z$-dependent, power-law extrapolation. The top panel of Fig.~\ref{fig:f_star_sc} compares these prescriptions at $z\sim8$.

Because the behavior at low luminosities is so important, we next consider recent observations from \citet{Atek_2015} that extend the measured luminosity function to $\approx -15.25$~mag using strong lensing from clusters.  We extend our abundance matching procedure at $z=7$ (the only redshift for which \citealt{Atek_2015} have substantial data) to this limiting magnitude using the Schechter parameters determined by that group. The resulting $f_{\star}$ is projected on top of our proposed extrapolations in the bottom panel of Fig.~\ref{fig:f_star_sc}, which indicates that $f_{\star}$ might scale as $M^{0.5}$ or slightly shallower in low-mass halos (Model~II). Here we only use the error in the faint-end slope $\alpha$ quoted by \citet{Atek_2015} in the match. 

\begin{figure*}
 \includegraphics[width=0.9\textwidth]{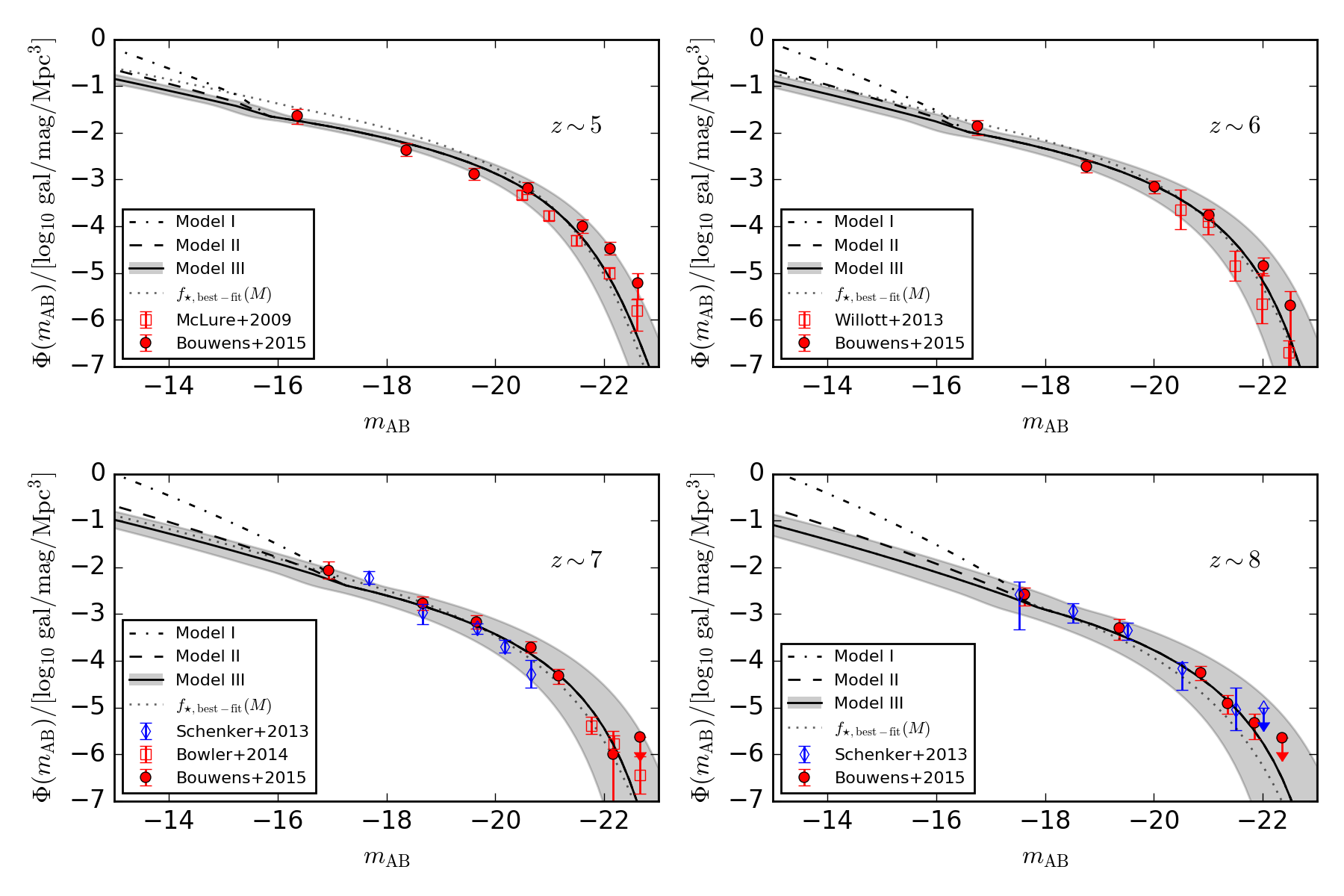}
 \caption{Comparison of observed UV luminosity functions at $z\sim5-8$ and those predicted by equation~(\ref{eq:fit_f_mz}). Here we do \emph{not} include star formation missing from UV observations because of dust obscuration. 1-$\sigma$ confidence intervals are shown by the shaded regions.}
 \label{fig:LF_CHECK}
\end{figure*}

Fig.~\ref{fig:LF_CHECK}  compares our parametric model to the observed UV luminosity functions at $5<z<8$. Note that when generating these model luminosity functions, we do not include star formation missing from UV observations because of dust obscuration, so these figures do \emph{not} show the total star formation rate density in the Universe.  The corresponding best-fit coefficients of equation~(\ref{eq:fit_f_mz}) are $a_{0}=-249.36\pm50.21, a_{1}=59.32\pm12.21, a_{2}=6.88\pm1.88, a_{3}=-0.57\pm0.15, a_{4}=-4.72\pm0.99$ and $a_{5}=0.13\pm0.03$. Below, when presenting predictions for the luminosity function we follow the same procedure, but when presenting estimates of the global star formation history and reionization history we always account for dust. As expected, our model agrees very well with the available data, with the only significant difference a modest underprediction of the bright end of the luminosity function at $z \sim 5$, at least according to the \citet{Bouwens_LF15} measurements.

\subsection{Comparison with previous work}

Abundance matching is a widely-used technique at both low and high redshifts, so here we compare our results to similar studies. At lower redshifts, the declining star formation efficiency to lower-mass halos is typically attributed to stellar feedback, either through winds generated by supernovae or by radiation pressure (see \citealt{L_F_2013BOOK} and references therein). The detailed effects of such feedback processes are not well-understood, but crudely they should produce a scaling $f_{\star} \propto M^\chi$, with $\chi \sim 1/3$--$2/3$, depending on whether the wind momentum or energy is the dominant factor. At $z\approx0$, \citet{Behroozi_LACK} find that, on average, $f_{\star}$ is proportional to $M_{h}^{2/3}$ below the characteristic halo mass at which $f_{\star}$ peaks. The mass scaling suggested by the \citet{Atek_2015} data is also consistent with this range.

\begin{figure}
\includegraphics[width=0.5\textwidth]{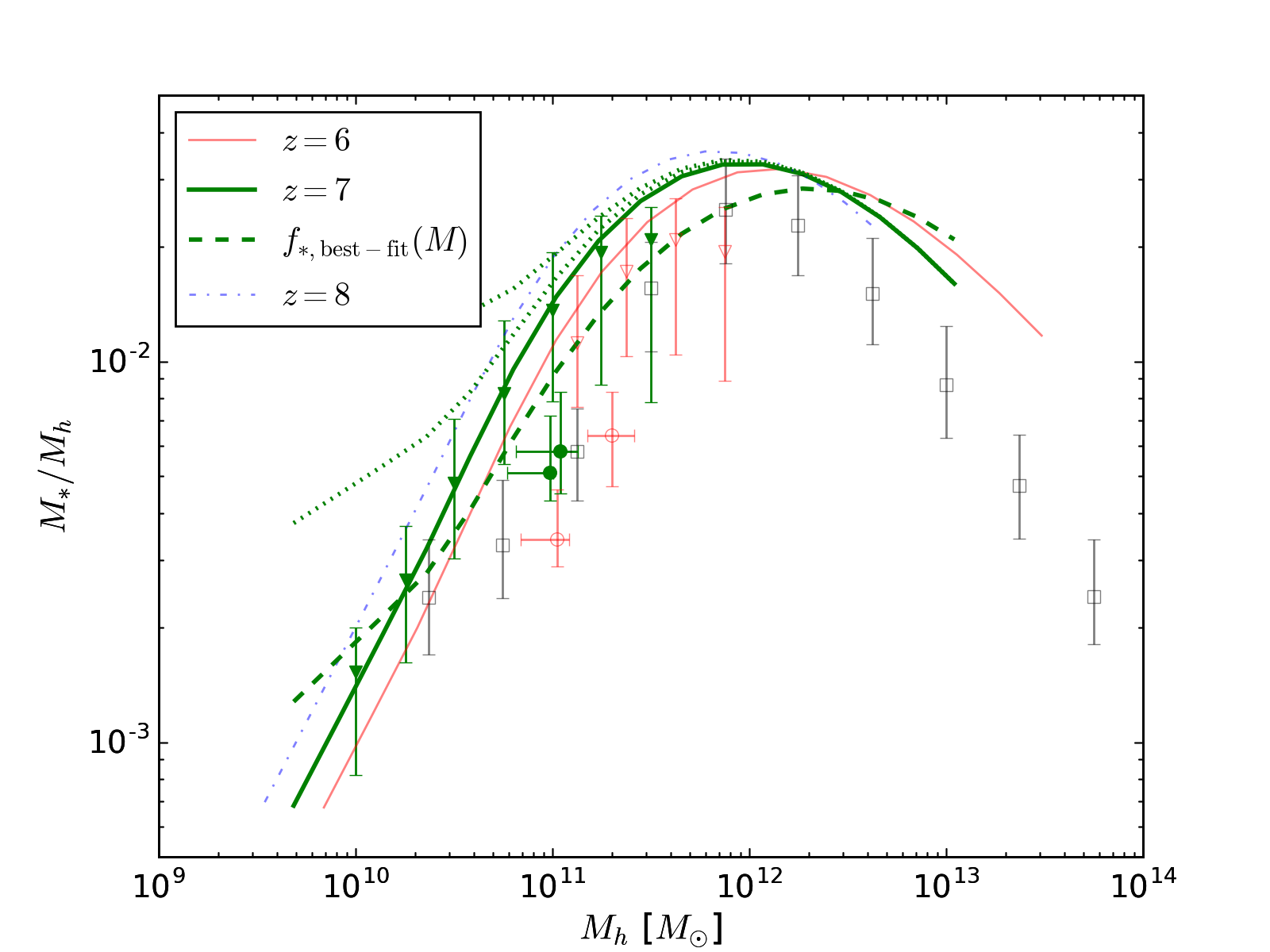}
\caption{Derived evolution of the ratio of stellar mass to halo mass at $z\sim6, 7, 8$ using the our model of continuous star formation, compared to the SMHM relations determined at $z\sim7$ (solid triangles, \citealt{B_S_2015}; solid circles, \citealt{Harikane_2015}) and those at $z\sim6$ (empty triangles, \citealt{B_S_2015}; empty circles, \citealt{Harikane_2015}) and in local universe (empty squares, \citealt{Behroozi_2013}). Solid lines represent Model III, whereas dashed and two dotted lines represent our $z$-independent fit, Model~II, and Model~I, respectively (bottom to top). Stellar masses are determined by tracing the continuous gas accretion of halos with $M_{h}>M_{\rmn{min}}$.}
\label{fig:SMHM}
\end{figure}

Much of the work on abundance matching, especially at lower redshift, focuses on the stellar mass to halo mass (SMHM) relation. We can calculate this ratio by integrating our star formation rates from the time halos reach $M_{\rm min}$, assuming a continuous star formation history for each halo. We compare to other work in Fig.~\ref{fig:SMHM}. As anticipated, the general shape of the SMHM ratio follows that of $f_{\star}(M,z)$, reaching the peak at approximately $10^{11.5-12.0}M_{\odot}$ at redshift $6<z<8$. This is also qualitatively consistent with the steadily decreasing average masses of $M_{\rmn{UV}}=-21$ galaxies, $\log\, (M_{h}/M_{\odot})=11.7$, 11.6, and 11.4 at $z = 5$, 6, and 7, respectively, found by \cite{Finkelstein_SBF} through direct abundance matching to the observed UV luminosity function (i.e. without a dust correction). Comparison with local measurements suggests that the peak halo mass might not have evolved significantly since $z\sim8$, whereas the peak value itself might have increased slightly with cosmic time, although with large uncertainties. 

\begin{figure*}
\includegraphics[width=0.9\textwidth]{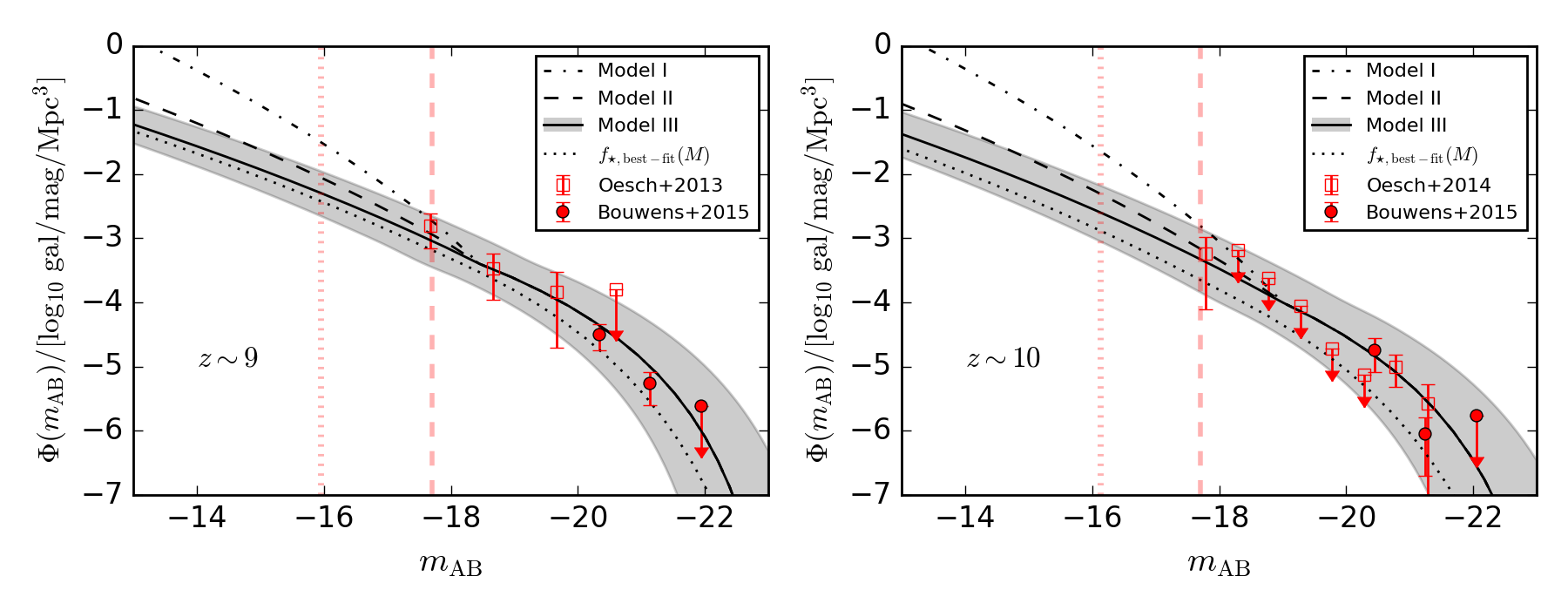}
\caption{
Comparison of predicted and observed luminosity functions at $z\sim9$ (\emph{top panel}) and $z\sim10$ (\emph{bottom panel}). The 1--$\sigma$ uncertainty level is shown for Model III. The vertical dashed and dotted lines show the UDF12 observational limit and the limiting magnitude of JWST, respectively. The latter assumes a $10^{6}$~s integration and a signal-to-noise ratio of 10. Note that the luminosity functions shown are predicted by our model of $f_{\star}$ without correcting for dust attenuation, which is expected to be trivial at $z>8$. 
}
\label{fig:LF_910}
\end{figure*}

We also compare our derived evolution of the SMHM ratio with that identified by \citet{B_S_2015} (see also \citealt{Behroozi_2013}) and \citet{Harikane_2015}. At $z>6$, we find a similar trend of increasing SMHM ratio with time at $M_{h}\sim10^{11}M_{\odot}$, although the increase is less rapid than the evolution inferred by \citet{Harikane_2015}, who used clustering observations with a modeled halo occupation distribution. 

Note that the difference between the SMHM ratios originally measured by \citealt{Behroozi_2013} and \citet{Harikane_2015} is mostly attributed to the distinct cosmologies and input models used (\citealt{B_S_2015} includes a redshift-dependent stellar mass correction relative to observed masses in order to fit data over a wide range of redshifts simultaneously), each contributing $\sim0.2$ dex of difference. The difference might also be associated with our simple continuous star formation histories (as opposed to the dark matter merger trees used by \citealt{B_S_2015}, though they argue that the merger channel is not a significant contributor at these redshifts). Accurate determination of the SMHM relation also relies on appropriate treatments of different feedback mechanisms, as demonstrated by recent simulations of dwarf galaxies (e.g. \citealt{Wheeler_2015}) which suggest a lower SMHM ratio than \cite{Behroozi_2013} and \cite{B_S_2015}. Nonetheless, given the qualitative similarity between the two independent methods and the large uncertainties associated with both of them, we do not pursue the explanation further. Moreover, following our modeled evolution of $f_{\star}(M,z)$, we find a nearly constant mass-to-light ratio $\rmn{d \, log} \,M / \rmn{d \, log} \, L$ over $6<z<8$ consistent with that measured by \citet{BN_2014} using galaxy clustering. 

Recently, abundance matching and similar techniques have also been applied to the $z > 5$ galaxy population in order to understand high-$z$ galaxy formation, particularly in light of the recent CMB measurements \citep{Mashian_2016, Mason_2015, Visbal_2015}. Compared to these works, we take a more general and rigorous approach to a measurement of $f_\star$, allowing both mass and redshift dependence and allowing a more general extrapolation to the critical faint galaxy population. In contrast, \citet{Mashian_2016} averaged over a range of redshifts to obtain $f_{\star}(M)$, while \citet{Mason_2015} calibrated two-parameter halo growth model at $z \sim 5$. Meanwhile, \citet{Visbal_2015} derived a redshift-dependent $f_{\star}$ by averaging the star formation efficiency over halos of different masses at a single redshift and then matching the normalization of the global star formation rate density to the halo growth rate. They then consider two models, one in which they fix the shape of $f_\star(M)$ and another in which they take a mass-independent average. 

In all of these similar studies, the results are qualitatively similar to ours but differ in the details. \citet{Mason_2015} find a stellar mass-halo mass relation very similar to our $z$-independent result, though with a slightly lower normalization. \citet{Mashian_2016} find a slightly smaller characteristic mass and a weaker decline at the high-mass end. In both of these cases, the models' predictions for the bright end of the luminosity function at $z \sim 9$--10 are very similar to ours, but their redshift-independent models predict a significantly steeper decline in the overall star formation rate toward high redshift (see the next section). \citet{Visbal_2015} allow a duty cycle in the star formation rate within galaxies (to represent their assumption that mergers dominate star formation) and therefore find a peak star formation efficiency at $M \sim 10^{10.5} \, M_\odot$, well below our value. 

Relative to these previous studies, our results demonstrate the importance of allowing for redshift evolution in $f_\star$ and in constraining that evolution in the future. Redshift-independent models are likely oversimplified as recent observations of galaxies at $z>4$ have shown evidence for a non-trivial evolution of the SMHM ratio \citep{Harikane_2015, Rod_2016}. While our measurements show only tentative evidence for redshift evolution, given the limited range of the observations, our best-fit result has significant implications for the star formation history, as we will show in the following sections. Allowing for such evolution is also important because of the rapidly-evolving halo mass function and the expectation that the feedback mechanisms governing star formation at these epochs are themselves redshift-dependent.


\section[]{Predictions for High-Redshift Galaxy Observations}

Because our model for the star formation efficiency uses theoretical inputs (the halo mass function and accretion rates) that can easily be extrapolated to higher redshifts, it is straightforward to test our results against higher redshift data and make predictions for future observations. Fig.~\ref{fig:LF_910} shows the predicted luminosity functions at $z\sim9$ and $z\sim10$ in comparison with recent measurements (\citealt{Oesch_2013}; \citealt{Oesch_2014}; \citealt{Bouwens_2015_z910}b). To date, data on the faintest galaxies disfavor Model I, but the constraints are not strong. As the primary tool to study the evolution of luminosity function of high-redshift galaxies, JWST will reach a detection limit about 2 magnitudes fainter than the current one (-17.7 mag, excluding lensing) with $10^{6}$~s of integration (shown by the vertical dotted lines), thereby providing a much better understanding of how $f_{\star}(M,z)$ might evolve in low-mass halos. 

Another straightforward observational implication of our model is the evolution of star formation rate density, $\rho_{\rmn{SFR}}$, which we estimate based on the simple formalism given by equation~(\ref{eq:def_f}) and our three extrapolations.\footnote{Note that the evolution of $f_{\star}$ is given by equation~(\ref{eq:fit_f_mz}) only at $M_{h} > 2 \times 10^{10}M_{\odot}$. Extrapolation is required below this mass threshold.} Fig.~\ref{fig:SFRD} shows $\rho_{\rmn{SFR}}$, and the equivalent UV luminosity density, predicted by our baseline models assuming $M_{\rmn{lim}}=-17.7\ \rmn{mag}$ and $\mathcal{K}_{\rmn{UV,1500}} = 1.15 \times 10^{-28} ~\rmn{M_{\odot}}~\rmn{yr}^{-1}/~\rmn{ergs}~\rmn{s}^{-1}~\rmn{Hz}^{-1}$. The solid and dashed curves in red represent $\rho_{\rmn{SFR}}$'s under Model III with and without a dust correction (at these bright limiting luminosities, all three of our models are identical except at very high redshifts). The dust correction becomes increasingly less important between $6<z<8$, due of the decreasing abundance of dusty massive galaxies at higher redshifts. The star formation rate density at $z\sim9$--$10$ predicted by our dust-corrected model is in good agreement with those recently determined from observations (green circles, \citealt{McLeod_2015}; orange squares, CANDELS/GOODS/HUDF, \citealt{Oesch_2014}; blue triangles, CLASH cluster searches, \citealt{Zheng_2012}, \citealt{Coe_2013}). The empty blue square shows the integrated star formation rate from \citet{Atek_2015} at $z=7$, which is consistent with all of our models.

There has been some controversy in the literature about the shape of the SFRD at high redshifts (e.g., \citealt{Robertson_HUDF}; \citealt{Oesch_2015}), reflected by the wide range in estimated star formation rates at higher redshifts shown by the data points in Fig.~\ref{fig:SFRD}. 
Model III implies a very slow steepening of $\rho_{\rmn{SFR}}$ with redshift, whereas Models~I and II predict a gradually flattening $\rho_{\rmn{SFR}}$. There is no strong evidence for a sudden drop of the dust-corrected $\rho_{\rmn{SFR}}$ and the overall trend is closer to the linear relation suggested by \cite{McLeod_2015}, although the exact slope depends on whether or not a dust correction is included. However, forced power-law fits to our dust-corrected model over $5<z<8$ and $8<z<12$ yield $\rho_{\rmn{SFR}}\propto(1+z)^{-6.0}$ and $\rho_{\rmn{SFR}}\propto(1+z)^{-9.0}$, respectively, comparable to scenarios favoring rapid evolution at $z\ga8$ \citep{Oesch_2015, Mashian_2016}. This apparent break is more a result of the fitting method than of any feature in our model, which evolves very smoothly in the context of continuous gas accretion. 

We note that the rapidly steepening $\rho_{\rmn{SFR}}$ predicted by various models \citep{Mashian_2016, Mason_2015} might be associated with the assumption of $z$-independent $f_{\star}$. As shown by the black dashed curve, an $f_{\star}$ fixed in time predicts substantially steeper evolution of SFRD with redshift. Interestingly, the apparent rate depends strongly on whether a dust correction is included, as that is larger for more massive halos and decreases quickly in importance at higher redshifts. Moreover, the extrapolation is very sensitive to assumptions about the behavior of low-mass galaxies. As indicated by the contrasting case of Model I (shown by the uppermost solid curve), different ways of extrapolating $f_{\star}$ leave much 
uncertainty in the $\rho_{\rmn{SFR}}$ evolution at $z>8$, calling for more detailed investigation by future observations. 

We also show, as a contrasting case, the cosmic $\rho_{\rmn{SFR}}$ predicted by the maximum-likelihood fit of \cite{Robertson_2015}, which extrapolates the observed $\rho_{\rmn{SFR}}$ evolution to a limiting luminosity of $L>0.001L_{*}$ (black, dotted curve), and forces smooth evolution from $z \sim 4$ via a double power-law model.  We compare with the prediction of Model II assuming the same limit (cyan, dotted curve). While the two approaches are remarkably similar considering the extrapolations involved, our model has significantly steeper redshift evolution than the purely empirical fit. This occurs because the low-mass halos (which become increasingly dominant at higher redshifts) are inefficient at producing stars in Model II. In the context of our approach, the Robertson et al. fit therefore implicitly assumes that low-mass galaxies become even more efficient at star formation toward higher redshift, demonstrating the importance of relating extrapolations to physical quantities in order to understand their systematic uncertainties. 

Finally, we present our predictions for future deep sky surveys that will be conducted by the \textit{James Webb Space Telescope} (JWST). For observations of the UV continuum, the flux density (or magnitude) can be acquired from the intensity of spatially unresolved objects. Knowing the limiting magnitude consequently allows us to estimate the integration time required to observe the majority of light from the faintest galaxies, as well as to calculate the galaxy number counts per unit area surveyed at a given redshift, which can be expressed as \citep{Pawlik_2011}
\begin{equation}
\frac{\mathrm d N}{\mathrm d \Omega}(>z) = \int_{z}^{\infty} \mathrm d z' \frac{\mathrm d V}{\mathrm d z' \mathrm d \Omega} \int_{L_{\rmn{min}}}^{\infty} \mathrm d L \frac{\mathrm d \phi(L,z')}{\mathrm d L}, 
\end{equation}
where ${\mathrm d V}/{\mathrm d z \mathrm d \Omega} = c d_{L}^{2}/(1+z)^{2}| {\mathrm d t} / {\mathrm d z}|$ is the comoving volume element per unit solid angle and redshift, and $L_{\rmn{min}}$ is the minimum observable UV luminosity. 

\begin{figure}
 \includegraphics[width=0.5\textwidth]{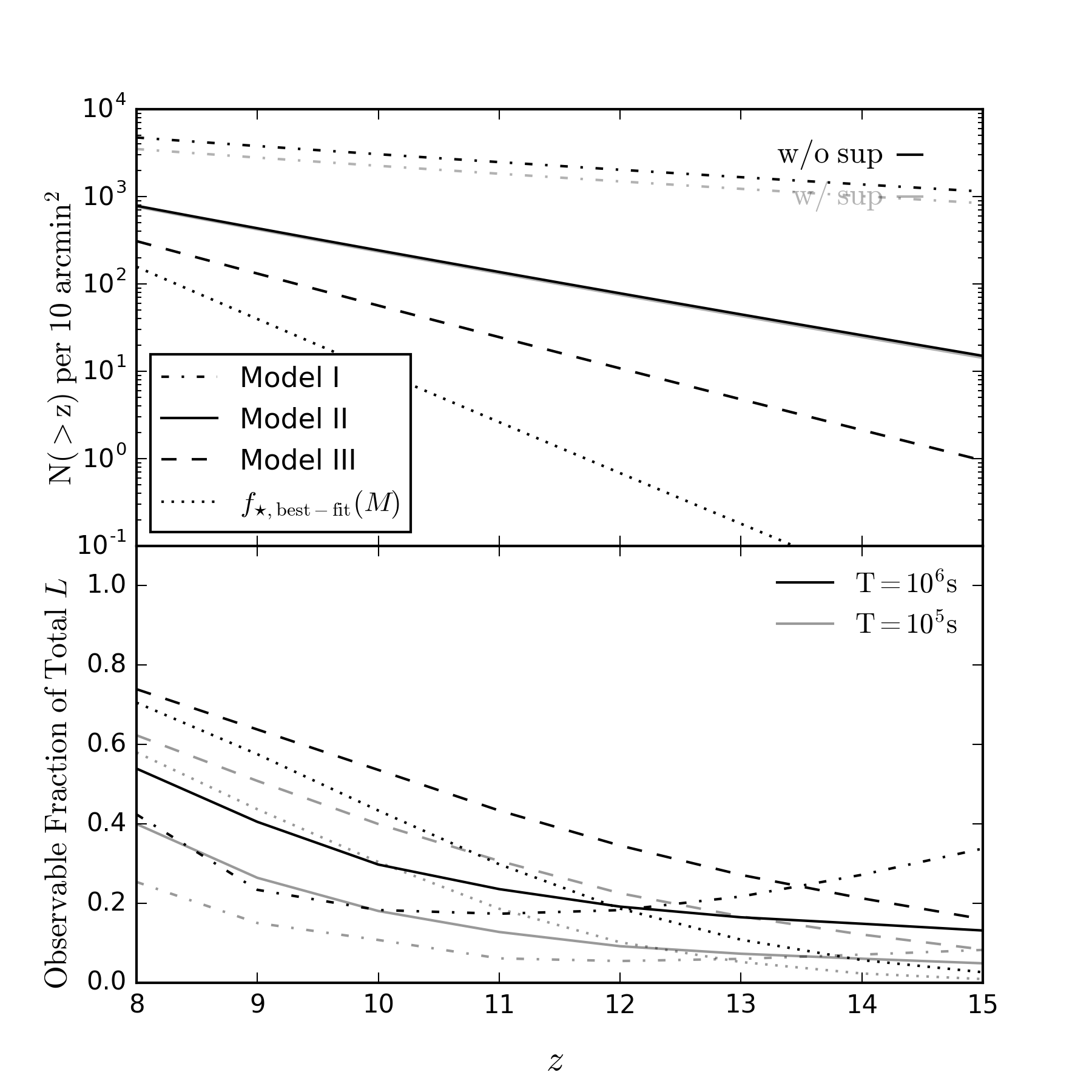}
 \caption{
\emph{Top:} Predicted galaxy number counts per $10\ \rmn{arcmin}^{2}$ for JWST assuming an integration time of $10^{6}$ seconds and 
requiring a signal-to-noise ratio of 10 for a detection. The three black curves show the different models for extrapolating to small halo masses. 
The grey curves represent cases which include photoionization suppression (corresponding to the violet curves in Fig.~\ref{fig:w_advanced_model}, respectively). \emph{Bottom:} Fractions of 
the total luminosity at a given redshift that can be observed by JWST (in the presence of photoionization suppression) with an integration time of $10^{5}$~s and $10^{6}$~s, respetively. The minimum halo mass is evaluated at the \protect\cite{Okamoto_2008} limit. 
}
 \label{fig:JWST_predictions}
\end{figure}

As shown by the vertical dotted curves
in Fig.~\ref{fig:LF_910}, JWST can significantly improve the determination of the faint-end slope and $f_{\star}$ evolution in faint galaxies, which helps better constrain star formation in the most primitive objects. Fig.~\ref{fig:JWST_predictions} shows the predicted galaxy number counts per 10 $\rmn{arcmin^{2}}$ and the fraction of the total luminosity density at a given redshift accessible to JWST for observations of $10^5$~s and $10^6$~s (grey and black lines, respectively). From the upper panel, Model I predicts many more observable galaxies for JWST than the other two models, in which $f_{\star}$ drops rapidly with decreasing halo mass. JWST will only detect substantial numbers of galaxies at $z>12$ in deep integrations if Model I is correct, but it is already in tension with observations at $z \sim 9$--10 (and with constraints on reionization, as we will see below). 
For an integration time of $10^{6}$~s, JWST is expected to observe roughly 10 UV-bright galaxy per 10 $\rmn{arcmin^{2}}$ beyond $z\sim15$ in Model~II. On the other hand, a typical $10^{6}$ second integration will allow JWST to observe more than half of the total UV continuum luminosity from star-forming galaxies at $z\sim8$ for Model~II and $z\sim11$ for Model~III. The apparent flattening (and even increase) of the observable fraction with redshift in Model~I is due to the increasing fraction of UV luminosity from faint galaxies, as predicted by the maximally extrapolated $f_{\star}$, rather than an actual increase of galaxy population or total UV luminosity. Therefore, assuming a star formation efficiency as described by Model~II or III, JWST is expected to determine the cosmic star formation rate between $8<z<11$ to better than 50\% and thereby provide much stronger constraints on how faint galaxies might have affected reionization. 

\section{Modeling Cosmic Reionization}

To this point, we have focused on direct interpretations of galaxy observations. These are limited in that they cannot yet probe the faint galaxy population at $z < 10$, and Fig.~\ref{fig:JWST_predictions} shows that the faint end at $z>10$ will likely remain hidden even after JWST begins observations. We next turn to combining our galaxy model with constraints on the reionization history. Specifically, the IGM ionization state is mostly determined by the emissivity of faint galaxies and is closely related to $M_{\rmn{min}}$ and $f_{\star}$. Thus it is helpful to compare models of reionization constructed from our inferred star formation history to various existing observational constraints. 

Meanwhile, adding these explicit probes of the faint galaxy population allows us to explore sensibly the ways in which new physics (beyond those relevant to lower redshift galaxies) may affect the star formation history. In this section, we shall introduce toy models for two such effects, the suppression of faint galaxies due to photo-heating during reionization and the introduction of Population~III stars. Our models will be highly simplified but will allow us to understand the trends imprinted by these physical processes.

The ionization state of intergalactic hydrogen can be measured by the globally-average ionized fraction, $x_{i}$, whose net rate of change is a balance between ionization and recombination,
\begin{equation}
\frac{\mathrm d x_{i}}{\mathrm d t} = \frac{\mathrm d (\zeta f_{\rmn{coll}})}{\mathrm d t} - \alpha_{B}(T_{e})\frac{C_{\rmn{H_{II}}}}{a^{3}}n_{\rmn{H}}^{0}x_{i}, 
\label{eq:reion_eqn}
\end{equation}
where $\zeta=A_{\rmn{He}}f_{\star}f_{\rmn{esc}}f_{\gamma}$ is the overall ionizing efficiency, a product of the correction factor for single ionization of helium, $A_{\rmn{He}}=4/(4-3Y_{p})=1.22$, the star formation efficiency, $f_{\star}$, the escape fraction of ionizing photons, $f_{\rmn{esc}}$, and the mean number of ionizing photons produced per stellar baryon, $f_{\gamma}$. In the second term, $\alpha_{B}(T_{e}) = 2.6 \times 10^{-13} (T_{e}/10^{4}\rmn{K})^{0.76} \rmn{{cm}^{3}{s}^{-1}}$ is the coefficient for case B recombination\footnote{In the early universe when ionizations were distributed fairly uniformly throughout the IGM, ionizing photons regenerated by recombinations to the ground state would not be re-absorbed locally in dense clumps but ionize other IGM atoms. Thus case B recombination is reasonable, though toward the end of the process recombinations in dense systems become important \citep{F_O_2005}. However, this is largely degenerate with uncertainty in the evolution of $C_{\rm H_{II}}$.} at electron temperature $T_{e}$, $C_{\rmn{H_{II}}} \equiv \left<n_{e}^{2}\right>/\bar{n}_{e}^{2}$ is the volume-averaged clumping factor of the IGM, and $n_{\rmn{H}}^{0}$ is the number density of hydrogen nuclei in local universe. In our calculations, we take $C_{\rmn{H_{II}}}=3$, which is a reasonable approximation during cosmic reionization (\citealt{Shull_2012}; \citealt{Pawlik_2015}) and $T_{e}=10^{4}\rmn{K}$. 

The value of $f_{\gamma}$ varies with the stellar population and thus depends on the IMF and metallicity. We take $f_{\gamma, \rmn{II}}\approx4,000$ for Population II stars, assuming a Salpeter IMF and $Z_{*}=0.05 Z_{\odot}$. In our baseline model, we assume all halos above $M_{\rmn{min}}$ form stars in a similar way so that
\begin{equation}
\frac{\mathrm d (\zeta f_{\rmn{coll}})}{\mathrm d t} = A_{\rmn{He}}f_{\gamma}f_{\rmn{esc}} \int_{M_{\rmn{min}}}^{\infty} f_{\star}\dot{M}_{\rmn{acc}}(M,z)\frac{\Omega_{\rmn{m}}}{\Omega_{\rmn{b}}} \frac{n(M)}{\bar{\rho}_{m}} \mathrm d M. 
\label{eq:baseline_model}
\end{equation}
Note that we use our estimate for $f_\star(M,z)$ in all of our calculations rather than imposing a value drawn from theoretical considerations. The exact value of $M_{\rmn{min}}$ is not known, so we leave it as a free parameter in our models, scaling to the value provided by Okamoto's criterion. It is especially important in Model I, where the star formation efficiency remains high inside small halos.

\subsection{Suppression of faint galaxies via photo-heating}

Simulations suggest that star formation in ``photo-sensitive'' halos less massive than a few times $10^{9}M_{\odot}$ will be suppressed by photo-heating as the IGM becoming reionized (\citealt{T_W_1996}; \citealt{Gnedin_2000}; \citealt{Finlator_2011}; \citealt{N_M_2014}). We therefore also consider a more sophisticated model, similar to those discussed by \cite{Alvarez_2012} and \cite{Visbal_2015}, in which we set a cut-off mass, $M_{\rmn{crit}}$, to distinguish halos in which star formation was gradually suppressed as cosmic reionization proceeded. As a simple model, we assume that halos with $M<M_{\rm crit}$ cannot form stars inside regions that have already been ionized. For simplicity, we ignore the relative bias of these halos and the ionized regions and we do not impose any delay on the suppression effect for the same reason. Consequently, $x_{i}$ evolves following
\begin{equation}
\begin{aligned}
\frac{\mathrm d (\zeta f_{\rmn{coll}})}{\mathrm d t} =\ & A_{\rm He} f_{\gamma} \left(\int_{M_{\rmn{min}}}^{M_{\rmn{crit}}} (1-x_{i}) + \int_{M_{\rmn{crit}}}^{\infty}\right) \\
& f_{\rmn{esc}} f_\star \dot{M}_{\rmn{acc}}(M,z)\frac{\Omega_{\rmn{m}}}{\Omega_{\rmn{b}}} \eta(z) \frac{n(M)}{\bar{\rho}_{m}}\mathrm d M. 
\label{eq:suppress_model}
\end{aligned}
\end{equation}

\subsection{Population~III stars in minihalos}

Another often-discussed complication to the early star formation history is the presence of Population~III stars. Their contribution is nearly completely unconstrained at present (though see \citealt{Visbal_2015} for weak constraints based on methods similar to our own). We therefore simply take a toy model of these objects in order to understand their implications for the reionization process.  In particular, we take $f_{\gamma, \rmn{III}}\approx40,000$ for Population III stars \citep{Bromm_KL_2001}. This value assumes that Pop~III stars are very massive and hence efficient ionizers. We further assume that they \emph{only} formed in halos where molecular cooling is efficient ($3000 \, \rmn{K}<T_{\rmn{vir}}<10^{4} \, \rmn{K}$), or halos of mass between $M_{\rmn{m}} \equiv M(T_{\rmn{vir}}=3000 \, \rmn{K})$ and $M_{\rmn{min}}$, with an arbitrary constant efficiency that is limited by the observed CMB optical depth. In order to model the transition from Pop~III to Pop~II star formation, we assume that a fraction 
\begin{equation}
\epsilon_{\rmn{III}}(z)=\frac{z-z_{\rmn{end}}}{(z_{\rmn{start}}-z_{\rmn{end}})} \hspace{0.3cm} \rmn{for}\ z_{\rmn{end}}<z<z_{\rmn{start}}
\label{eq:PopIII_frac}
\end{equation}
of halos in the relevant mass range (and outside of ionized regions, where photoheating will suppress such sources) are able to form stars.
We take $z_{\rmn{start}}=30$ to be the fiducial value for the time when PopIII stars started to form and vary $z_{\rm end}$.

Because these halos are subject to photoheating suppression, when studying Pop~III stars we use the following model for the ionizing emissivity:
\begin{equation}
\begin{aligned}
\frac{\mathrm d (\zeta f_{\rmn{coll}})}{\mathrm d t} =\ & \left[f_{\gamma, \rmn{III}}\epsilon_{\rmn{III}}\int_{M_{\rmn{m}}}^{M_{\rmn{min}}} f_{\star,\rmn{m}}(1-x_{i})\right. \\
& + \left. f_{\gamma, \rmn{II}} \left(\int_{M_{\rmn{min}}}^{M_{\rmn{crit}}} f_{\star}(1-x_{i}) + \int_{M_{\rmn{crit}}}^{\infty}f_{\star} \right)\right] \\
& A_{\rmn{He}} f_{\rmn{esc}} \dot{M}_{\rmn{acc}}(M,z)\frac{\Omega_{\rmn{m}}}{\Omega_{\rmn{b}}} \frac{n(M)}{\bar{\rho}_{m}}\mathrm d M. 
\label{eq:advanced_model}
\end{aligned}
\end{equation}
Note that, over the range $M_{\rmn{m}}<M<M_{\rmn{min}}$, we let $f_{\star}=f_{\star,\rmn{m}}$ be a free parameter and do not adopt any form of extrapolation used for atomic cooling halos.


\begin{figure*}
\begin{subfigure}{.47\textwidth}
\centering
\includegraphics[width=.99\textwidth]{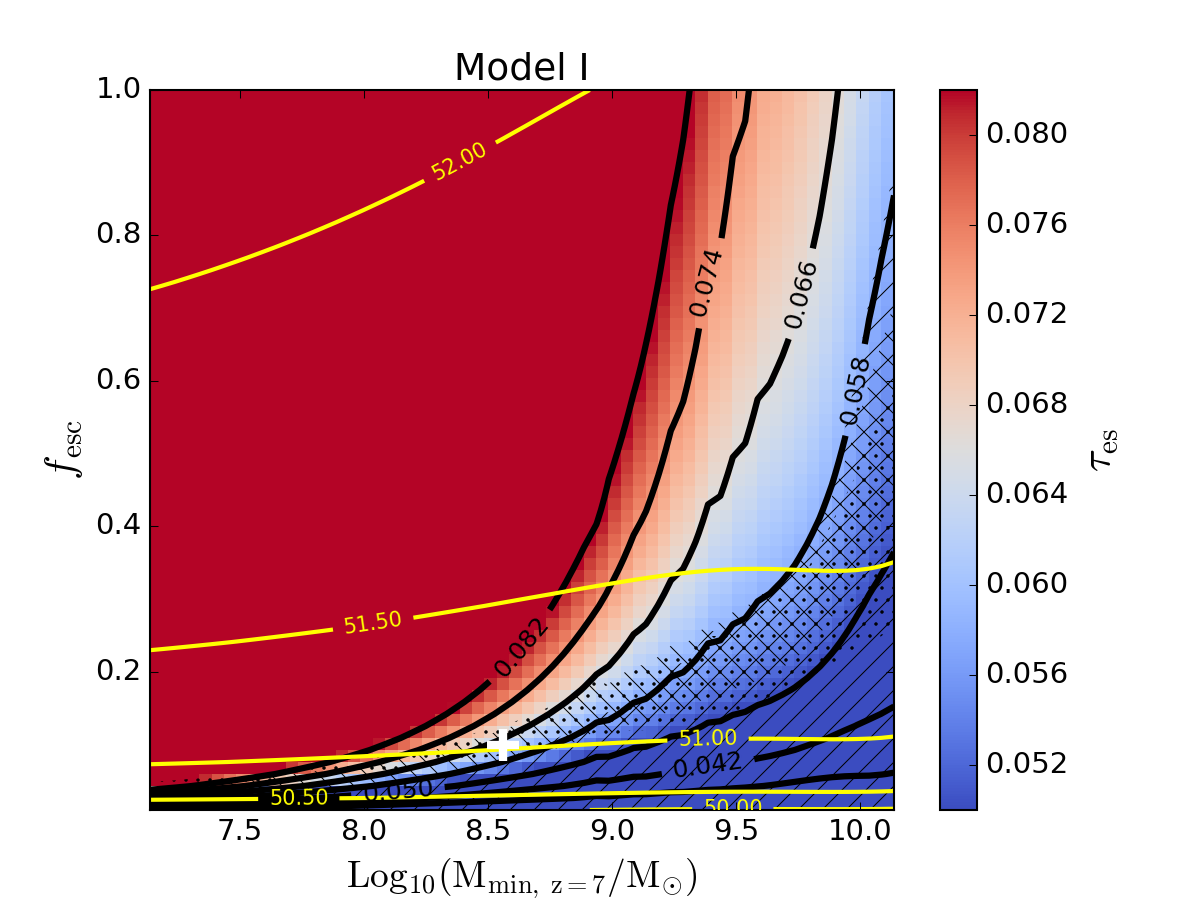}
\label{fig:Para_Space_UL}
\end{subfigure}
\begin{subfigure}{.47\textwidth}
\centering
\includegraphics[width=.99\textwidth]{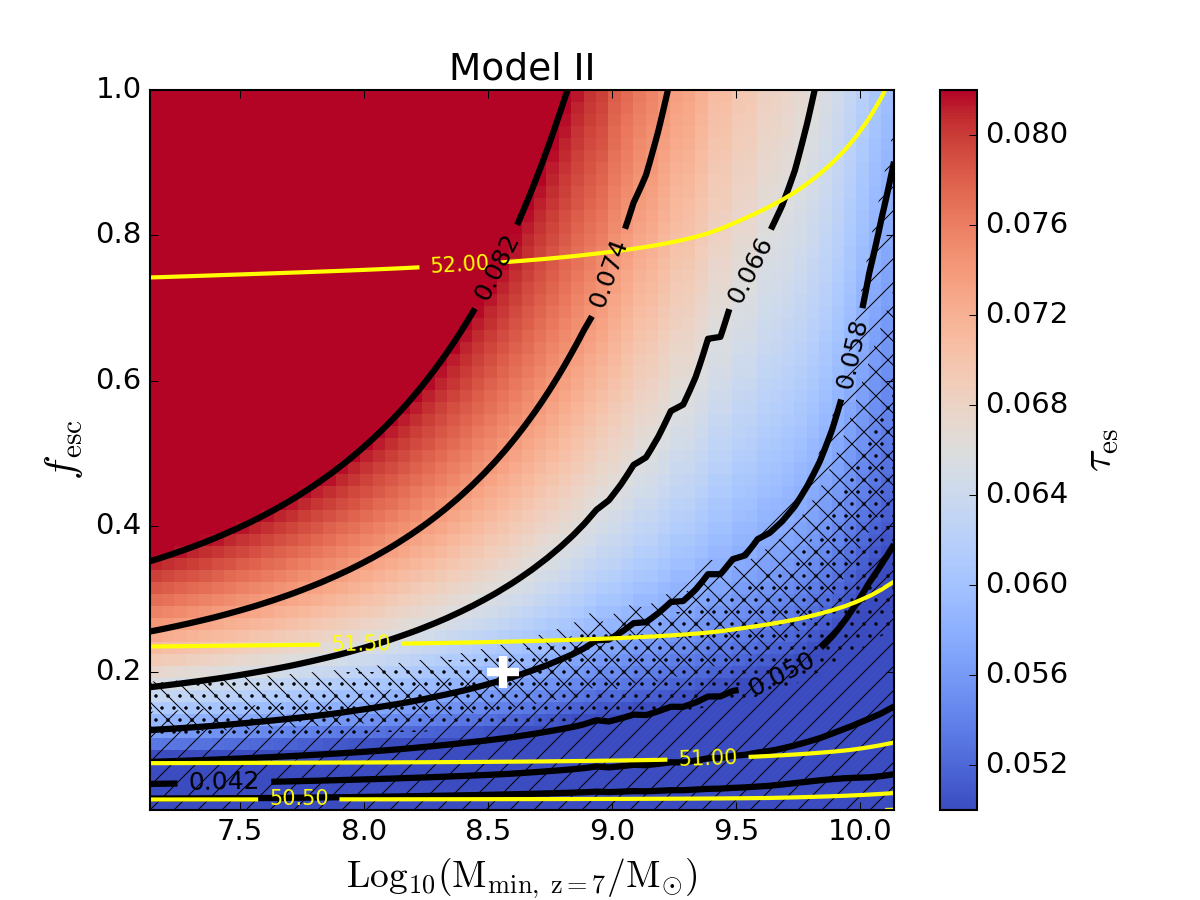}
\label{fig:Para_Space_HALF}
\end{subfigure}
\begin{subfigure}{.47\textwidth}
\centering
\includegraphics[width=.99\textwidth]{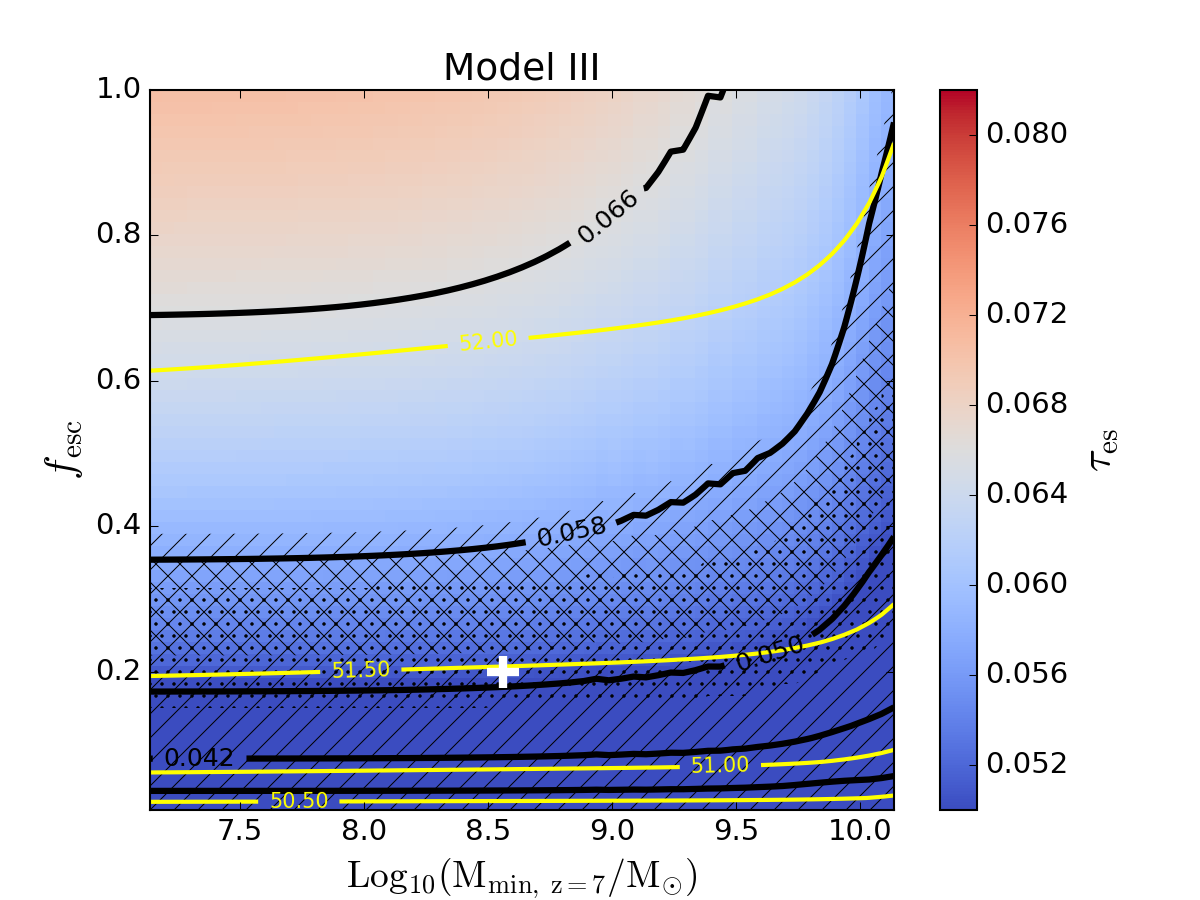}
\label{fig:Para_Space_EXT}
\end{subfigure}
\begin{subfigure}{.47\textwidth}
\centering
\includegraphics[width=.99\textwidth]{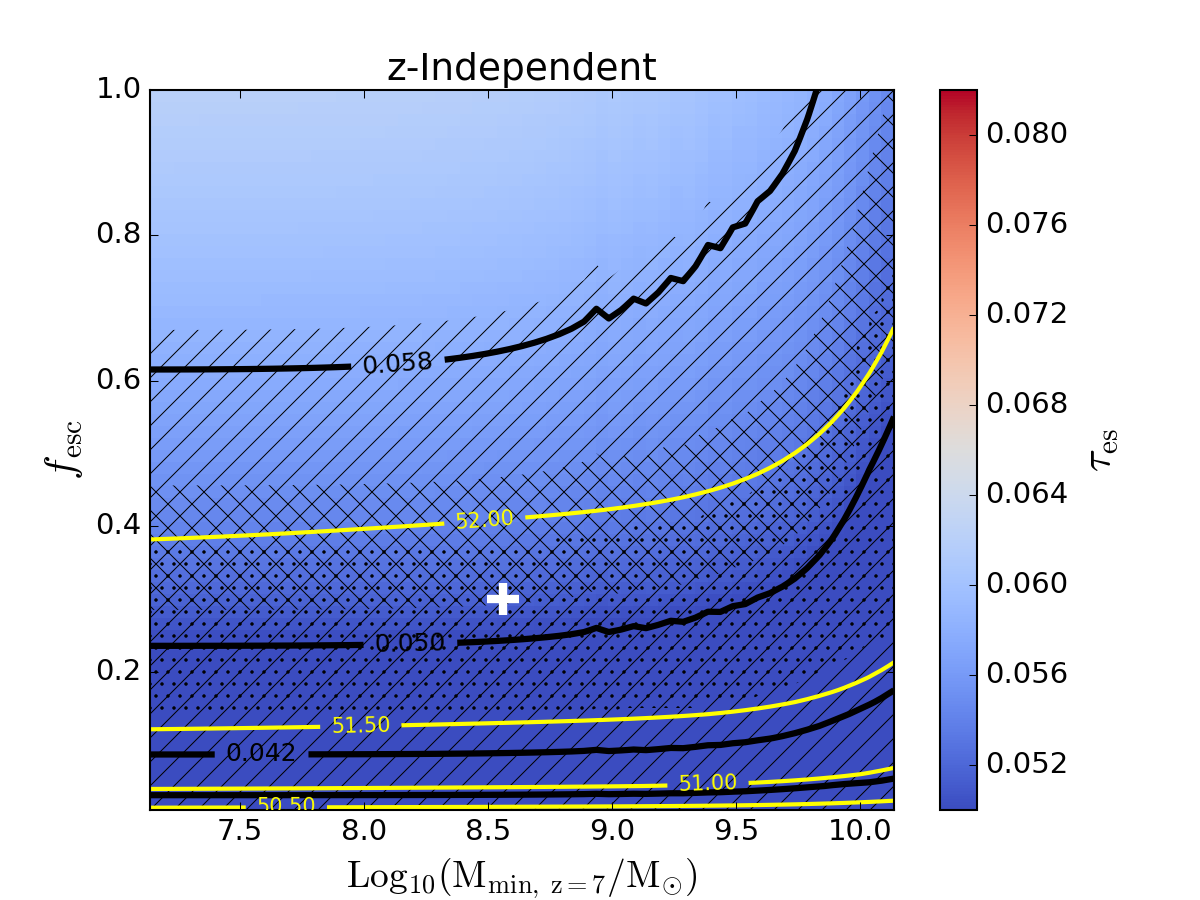}
\label{fig:Para_Space_ZI}
\end{subfigure}
\caption{Observationally constrained parameter space for the baseline model of reionization, assuming maximally (top-left panel), moderately (top-right panel), and minimally (bottom-left panel) extrapolated $f_{\star}$ (Models I, II, and III, respectively). The parameter space for the redshift-independent model is also shown for comparison (bottom-right panel). The plot is color-coded by the predicted Thomson scattering optical depth, with the thick black contours comparing to the value measured by the Planck satellite, $\tau_{\rmn{es}}=0.066 \pm 0.016$. Yellow curves show model predictions for the logarithm of the emissivity of ionizing photons, $\log_{10} (\dot{N}_{\rmn{ion}}/[\rmn{s^{-1}Mpc^{-3}}])$, evaluated at redshift $z=5$, which is measured to be approximately $51.0\pm0.5$ by \protect\cite{BB13}. Dot-hatched region represents the allowed parameter space for which $5.8 \leq z_{\rmn{ion}} \leq 6.6$ as inferred from the spectra of high-redshift QSOs, where $z_{\rmn{ion}}$ is the redshift at which reionization completed (i.e. $x_{i} \approx 1$). Portions of parameter space preferred by observational constraints on the evolution of ionization state, $x_{i}$, are shown by the regions with slash and back-slash hatch, corresponding to $x_{i}(z=7)=0.65\pm0.15$ and $x_{i}(z=8)\leq0.35$, respectively (\citealt{Stark_2010}; \citealt{Treu_2013}; \citealt{Faisst_2014}; \citealt{Pentericci_2014}; \citealt{Schenker_2014}; \citealt{Tilvi_2014}). White crosses indicate reasonable combinations of $f_{\rmn{esc}}$ and $M_{\rmn{min}}$ 
later used in Fig.~\ref{fig:w_advanced_model}.}
\label{fig:2D_color_plots}
\end{figure*}

\section[]{Joint Analysis of Galaxy Populations and the Reionization Observables}

Our models of galaxy growth and reionization have three key free parameters: $f_\star(M,z)$, $M_{\rm min}$, and $f_{\rm esc}$. In Section 2, we constrained the first, at least over a wide mass range, through observations. We will next use ancillary measurements of the integrated Thomson optical depth, global IGM neutrality, and ionizing background at $z<6$ to quantify the interdependencies and degeneracies between these parameters. 

For concreteness, we employ the baseline model described by equation~(\ref{eq:baseline_model}). For each model of $f_{\star}$, we limit the 2D parameter space formed by $M_{\rmn{min}}$ and $f_{\rmn{esc}}$. We allow $f_{\rmn{esc}}$ to vary between 0.01 and 1. Observations of star-forming galaxies at $z\sim1$--$3$ suggest a small average ionizing escape fraction from a few percent to smaller than 15\%, which is likely to increase towards fainter galaxies (or equivalently towards higher redshifts, e.g., \citealt{Iwata_2009}; \citealt{Siana_2010}; \citealt{Nestor_2013}; \citealt{Mostardi_2015}). In recognition of the redshift dependence of $M_{\rmn{min}}$, we model its variation by rescaling $M_{\rmn{min}}(z)$ as given by \cite{Okamoto_2008} by a factor of 0.1--10. For the convenience of discussion, the rescaled $M_{\rmn{min}}$ is always quoted at $z\sim7$. 

\begin{figure*}
 \includegraphics[width=0.99\textwidth]{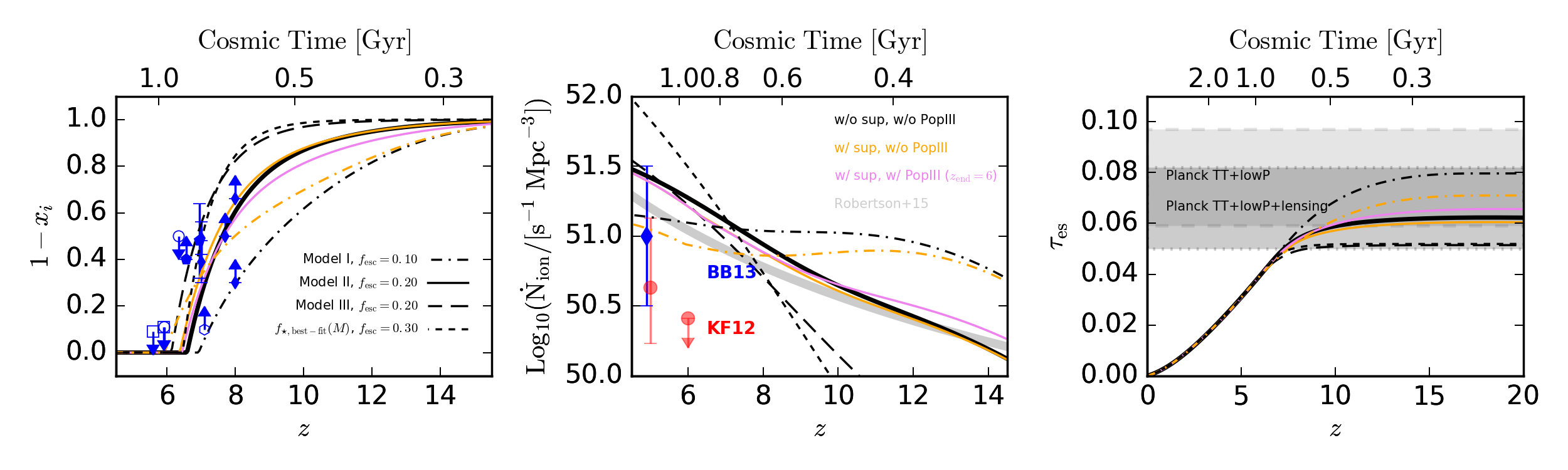}
 \caption{The IGM neutrality (left), emissivity of ionizing photons (center), and Thomson optical depth (right). The lines are the same in each panel, with the styles indicating Models I, II, III and the $z$-independent model and the colors indicating variations in the stellar populations and feedback prescriptions (the case with both photoionization suppression and PopIII stars is shown in violet for Model~II only). See the text for details on the model parameters. We compare the predictions to a number of observational constraints (see text) in each panel. Note that including photoionization suppression makes little difference in Model~III and the $z$-independent model and therefore the orange curves are only shown for Models~I and II.}
 \label{fig:w_advanced_model}
\end{figure*}

\subsection{The CMB optical depth}

We compare our models to three important measurements of the ionization history. Our first constraint is the Thomson scattering optical depth for CMB photons, which provides an integral measure of the reionization history. This is given by
\begin{equation}
\tau_{\rmn{es}} = \frac{3H_{0}\Omega_{b}c\sigma_{T}}{8\pi G m_{p}}\int_{0}^{\infty}\frac{x_{i}(z)(1+z)^{2}(1-Y_p+N_{\rmn{He}}(z)Y_p/4)}{\sqrt{\Omega_{m}(1+z)^3+1-\Omega_{m}}} \mathrm d z, 
\end{equation}
where $N_{\rmn{He}}=1$ $(z>3)$ or $2$ $(z<3)$ measures the degree of helium ionization.\footnote{Here we assume helium is completely reionized at $z \sim 3$, in agreement with current observations (\citealt{Fechner_2006}; \citealt{Shull_2010}; \citealt{S_S_2014}).} The CMB optical depth is useful because it is well-measured by recent experiments (Planck Collaboration 2015, though see below). However, it is an integral constraint over \emph{both} the faint galaxy population \emph{and} the overall reionization history, so it still allows a great deal of freedom in models.  Fig.~\ref{fig:2D_color_plots} is color-coded by $\tau_{\rm es}$ and quantifies the $M_{\rmn{min}}$--$f_{\rmn{esc}}$ degeneracy under Model I (top-left), II (top-right), III (bottom-left) and our $z$-independent model (bottom-right). The thick black contours are marked according to the measurement of $\tau_{\rmn{es}}=0.066\pm0.016$ reported by the \cite{Planck_Team} in units of $0.5\sigma$ (assuming gaussian statistics); dark regions are disfavored by those observations.

As expected, given a specific optical depth $\tau_{\rmn{es}}$, $f_{\rmn{esc}}$ and $M_{\rmn{min}}$ are positively related, and the relationship depends on the relative abundance of faint galaxies encoded by the extrapolation of $f_{\star}$. Under Model I, in which $f_{\star}$ is maximally extrapolated and therefore faint galaxies are abundant, $f_{\rmn{esc}}$ must grow rapidly with increasing $M_{\rmn{min}}$ to maintain a constant $\tau_{\rmn{es}}$, whereas under Models~II and III $f_{\rmn{esc}}$ is much less sensitive to $M_{\rmn{min}}$ due to 
their rapidly diminishing populations of faint galaxies. This suggests that, even though $M_{\rmn{min}}$ and $f_{\star}(M,z)$ enter our models as two independent parameters, observations place strong constraints on their joint properties \emph{in the context of a particular star formation model}. Inferences about the escape fraction also rely on the assumed extrapolation of $f_{\star}$. In Model I, $f_{\rmn{esc}}$ as small as 0.1 is large enough to reproduce the observed $\tau_{\rmn{es}}$ with a moderate $M_{\rmn{min}}$ comparable to our fiducial model. But in Models II and III, in which $f_{\star}$ declines rapidly with decreasing halo mass, $f_{\rmn{esc}} \ga 0.2$ is required to reach just the $1\sigma$ lower bound of Planck's $\tau_{\rmn{es}}$ at the same $M_{\rmn{min}}$ \citep{Robertson_HUDF, Robertson_2015}. 

If we assume no redshift dependence and take our best fit $f_\star$ to the aggregate data from $z=5$--8, it is very difficult to match the $\tau_{\rmn es}$ observations without appealing to $f_{\rmn{esc}} > 1$, which would in practice require changing the stellar populations to be much more efficient ionizers.  This is because our $z$-independent fit maintains $M_{\rm peak}$ at $M \sim 10^{12}$~M$_\odot$, far above the characteristic mass at very high redshifts. The shift of $M_{\rm peak}$ toward lower masses in our fiducial model significantly increases the contribution of high-$z$ galaxies to the total optical depth.

We emphasize that these conclusions are all in the context of our simple model. For example, the high values of $f_{\rm esc}$ we require (in comparison to measurements at lower redshifts) can be ameliorated by introducing a new population of ionizing sources at very high redshifts, which cannot be modeled properly by our extrapolations.  

Fig.~\ref{fig:w_advanced_model} illustrates the difficulty of matching the optical depth in our fiducial model. 
The right-hand panel shows the integrated optical depth, assuming $f_{\rm esc}=0.1,\,
0.2,$ and 0.2 for our three $z$-dependent models and $f_{\rm esc}=0.3$ for our $z$-independent model (marked by the white crosses in Fig.~\ref{fig:2D_color_plots}). Only Model I can easily reach the best-fit value from Planck, 
though all are within the $1\sigma$ lower bound.
We note that while the $\rmn{TT+lowP+lensing}$ dataset of Planck suggests a lower bound (1$\sigma$) of the CMB optical depth as low as $\sim0.05$ \citep{Planck_Team}, the $\rmn{TT+lowP}$ dataset itself indicates a higher optical depth $\tau=0.078\pm0.019$ (i.e., an earlier reionization). The difference between these likely roughly captures the level of systematic uncertainty in CMB measurements \citep{Planck_July}. As shown by Fig.~\ref{fig:w_advanced_model}, the higher optical depth can be reached by assuming more efficient star formation in low-mass halos (close to our Model~I) or allowing a higher escape fraction $>0.2$ in our fiducial model. While the former is permitted by the existing observational constraints, we will see that the latter poses some difficulties.

\subsection{Other Reionization Constraints}

We next ask whether other existing observational constraints on $x_i(z)$ further limit the allowed parameter space. These measures of the instantaneous neutral fraction provide different information than $\tau_{\rm es}$, because they help us distinguish the progress of different models across cosmic time. This also implies that they require modeling of the entire ionization history, which is provided by our galaxy evolution model.

The hatched regions stretching from the lower left corner in the panels of Fig.~\ref{fig:2D_color_plots} show parameter space permitted by various indirect measures of the neutral fraction at particular redshifts. 
We show the constraints on $x_i$ themselves as symbols in the left panel of Fig.~\ref{fig:w_advanced_model}. 
First, the dot-hatched region in Fig.~\ref{fig:2D_color_plots} shows models where reionization completes $5.8 \leq z_{\rmn{ion}} \leq 6.6$, as inferred from the spectra of high-redshift QSOs \citep{Fan_2006, Bolton_2011, McGreer_2015}; these are also shown by the solid diamonds and open squares in Fig.~\ref{fig:w_advanced_model}. The regions filled by slash and back-slash hatch mark model where the global ionized fraction are $x_{i, z\sim7}=0.65\pm0.15$ and $x_{i, z\sim8}\leq0.35$ at $z\sim7$ and $z\sim8$, respectively, in agreement with the measurements of declining Lyman alpha emission fraction of high-redshift galaxies (\citealt{Stark_2010}; \citealt{Treu_2013}; \citealt{Faisst_2014}; \citealt{Pentericci_2014}; \citealt{Schenker_2014}; \citealt{Tilvi_2014}; see also the solid pentagons in Fig.~\ref{fig:w_advanced_model}). 
The limits on the IGM neutrality inferred from the measurements of quasar near zones (open hexagon, \citealt{Mortlock_2011}) and the GRB damping wings (open circle, \citealt{Chornock_2013}) are also shown in the left panel of Fig.~\ref{fig:w_advanced_model}. We note that we regard all these measurements as model-dependent thanks to the difficulty in evaluating the systematic modeling uncertainties from which constraints on the neutral fraction are drawn.

The parameter space favored by these measurements roughly traces the $-1\sigma$ to $-0.5\sigma$ interval of $\tau_{\rmn{es}}$, suggesting a broad agreement between the ionizing efficiency required by the observed history of reionization and that actually supplied by star-forming galaxies. However, there is some tension between the baseline model and these observables, as shown by the left panel of Fig.~\ref{fig:w_advanced_model}. In particular, Model~I is in marginal disagreement with some of these measures, because it relies on faint galaxies that evolve slowly throughout this epoch. The others, with star formation centered on more massive halos, evolve more rapidly and satisfy all the existing constraints.

\subsection{The Ionizing Emissivity}

Finally, we compare our models to the Lyman-limit photon emissivity $\dot{N}_{\rmn{ion}}$, which can be measured from the Lyman-$\alpha$ forest at $z \la 5$ (\citealt{KF12}, KF12; \citealt{BB13}, BB13). In our model, this equals
\begin{equation}
\dot{N}_{\rmn{ion}} = \frac{\mathrm d (\zeta f_{\rmn{coll}})}{\mathrm d t} \bar{\rho}_{m} \frac{\Omega_{b}}{\Omega_{m}} \frac{1}{m_{p}},
\end{equation}
The total emissivity is useful as a constraint on the integrated population of galaxies at the tail end of reionization, though of course it is an imperfect measure because of potential redshift evolution. The yellow lines in Fig.~\ref{fig:2D_color_plots} show the inferred emissivity at $z=5$ in our models. These should be compared to the measurement from BB13 at $z \sim 5$, $\log_{10} (\dot{N}_{\rmn{ion}}/[\rmn{s^{-1}Mpc^{-3}}])=51.0\pm0.5$. 

While it is possible to match both $\tau_{\rm es}$ and constraints on $x_i$ without undue difficulty, further matching the ionizing emissivity at $z\sim5$, right after the completion of reionization, is challenging, particularly in Models~II and III. This tension can also be seen in the middle panel of Fig.~\ref{fig:w_advanced_model}, which shows the evolution of the ionizing emissivity. Here we show the BB13 measurement as well as that from KF12. The difference results from BB13's choice of a lower IGM temperature and more accurate treatments of the optical depth and cosmological radiative transfer effects. Therefore, we adopt BB13 as our fiducial limit on the ionizing emissivity at $z\sim5$. Models~II and III, which rely on massive galaxies to reionize the Universe, have a rapid increase in the ionizing emissivity that only marginally agrees with the observed values at $z \sim 5$, if they are calibrated to complete reionization at $z \sim 6$. This demonstrates a tension between the emissivity measurements and $\tau_{\rm es}$: increasing $f_{\rm esc}$ improves agreement with the latter but reduces agreement with the former. This is one motivation to explore additional physics in the following subsections. 
For comparison, we also present the emissivity derived with the SFRD given by \citet{Robertson_2015}, which provides a slightly better match to the observations while being largely consistent with our fiducial model. 

Note, however, that the low $f_{\rmn{esc}}$ required by Model I avoids over-predicting the post-reionization ionizing background. In particular, Model~I fits nicely with recent simulation results that suggest a low time-averaged $f_{\rmn{esc}}\la0.05$ with negligible mass and redshift dependence \citep{Ma_2015}. If Model I or a very gradual extrapolation of $f_{\star}$ is true, then $\dot{N}_{\rmn{ion}}$, together with other observational constraints, suggests that $f_{\rmn{esc}}$ might not exceed 0.1. In that case, the minimum halo mass could not differ from Okamoto's criterion by more than 
$\sim0.5$~dex. 


\section{Implications of Additional Source Physics}

Next, we discuss variations from our baseline model. The effects we will explore here are highly uncertain, from the perspectives of both modeling and observations. Thus we simply attempt to provide a qualitative understanding of their relative importance by contrasting them with our baseline model.

\subsection{Photoionization Suppression}

We begin by considering the effects of the suppression of galaxy formation from photoheating during reionization, as described by 
equation~(\ref{eq:suppress_model}) in Section 4.1. In the following discussion, we assume the critical halo mass $M_{\rmn{c}}$ below which star formation is ``photo-suppressible'' to be $5.0\times10^{9}M_{\odot}$, consistent with the mass scale for efficient photoionization feedback (\citealt{Gnedin_2000}; \citealt{Finlator_2011}), while all other parameters take the same values as in the fiducial model, unless otherwise stated. 

To see how photoionization heating might influence the reionization history, we compare our revised models to the baseline model given a specific combinations of $M_{\rmn{min,z=7}}=3.6 \times 10^{8}M_{\odot}$ and $f_{\rmn{esc}}=0.1$, 0.2, 0.2 for Model I, II, III, respectively (see the white crosses in Fig.~\ref{fig:2D_color_plots}), so that the predicted histories of reionization are in general agreement with the observational constraints, in particular the optical depth and the redshift of completion. 

The orange curves in Fig.~\ref{fig:w_advanced_model} compare Models I and II with photo-heating suppression to the fiducial model discussed previously (shown in black). Including photoionization suppression clearly diminishes the contribution 
to the total ionizing emissivity from faint galaxies, resulting in a delayed reionization. This effect is particularly 
significant for Model I (dot-dashed curves), which includes a large population of photo-suppressible galaxies hosted by halos below $2 \times 10^{10} M_{\odot}$, whereas in other models the effect of photoionization is almost indiscernible. In these models, the low-mass galaxies already have only a very small contribution to the ionizing emissivity, so their suppression has very little effect on the overall emissivity. Thus the tension between observed and predicted post-reionization emissivities under Model~III is not resolved by simply including the feedback from photoionization heating.  

This is one of the most important conclusions from our analysis: photoheating can only be a significant feedback mechanism if the shape of $f_\star(M) \times f_{\rm esc}$ differs \emph{qualitatively} from that at low redshifts, with much more efficient star formation in faint galaxies. While we cannot rule out such a scenario, it is another indication -- along with the high required $f_{\rm esc}$ values -- that galaxies may undergo substantial evolution at $z \ga 6$.

\begin{figure}
 \includegraphics[width=0.5\textwidth]{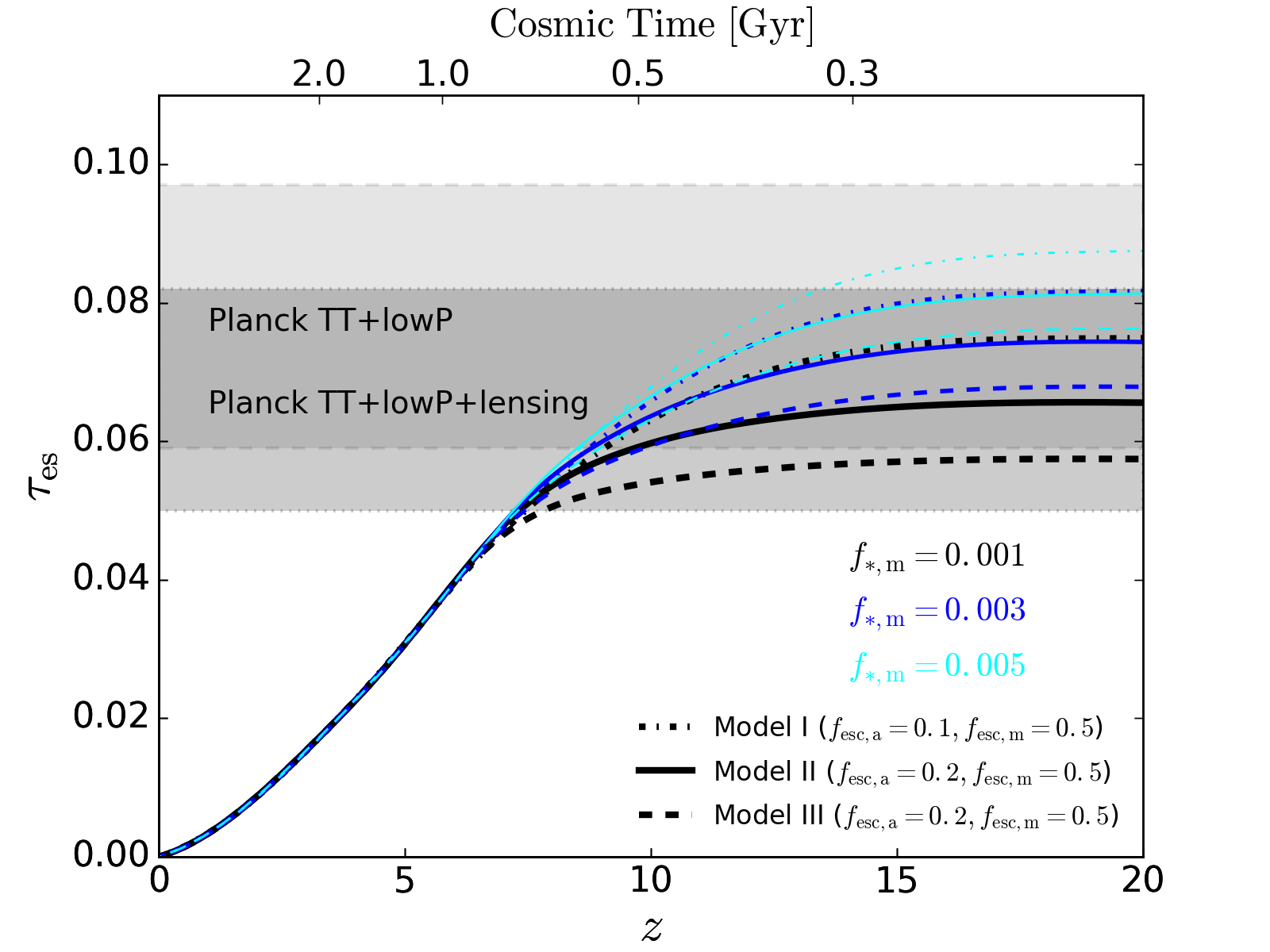}
 \caption{Optical depth to electron scattering as a function of Pop~III star formation efficiency $f_{*,\rmn{m}}$ of molecular cooling halos, in the presence of photoionization suppression and a cessation of Pop~III star formation at $z_{\rmn{end}}=6$. Black, blue, and cyan curves (with decreasing thickness)
take $f_{*,\rmn{m}}=0.001,\ 0.003\ \rmn{and}\ 0.005$, respectively. 
}
 \label{fig:tau_PopIII_comp}
\end{figure}

\subsection{Population III Stars}

Next, we present a crude analysis of the contribution from massive Pop~III stars based on the simple model described in equation~(\ref{eq:PopIII_frac}) and Section 4. We refer readers to \cite{Visbal_2015} for a more careful analysis of the implications for Pop~III stars in light of the recent $\tau_{\rm es}$ measurements from Planck. Here, we only consider the case where $z_{\rmn{end}}=6$, consistent with both models and numerical simulations of Pop~III star formation regulated by a variety of feedback mechanisms, which suggest Pop~III stars could have formed as early as $z\sim20-30$ and a small amount of them might still form until the end of reionization \citep{Scannapieco_2003, F_L_2005, Trenti_2009, Johnson_2013}. 

To set $f_{\rmn{*,m}}$, we consider in Fig.~\ref{fig:tau_PopIII_comp} the total contribution of this population to $\tau_{\rm es}$, 
assuming a fixed escape fraction $f_{\rmn{m, esc}}=0.5$ for Pop~III stars since effects like supernova blowout and weaker dust attenuation could allow more photons to escape these fragile halos. When $f_{\rmn{*,m}} = 5 \times 10^{-3}$, all models reach or exceed Planck's $1\sigma$ upper limit of $\tau_{\rmn{es}}$. 
We use this calculation to choose $f_{\rmn{*,m}} =10^{-3}$ as a plausible star formation efficiency in the minihalo population. It is also consistent with the results of \cite{Visbal_2015}, which suggest the formation efficiency of Pop~III stars is tightly constrained by Planck data. 

Now taking $f_{\rmn{*,m}} =10^{-3}$ and $z_{\rmn{end}}=6$, Fig.~\ref{fig:w_advanced_model} contrasts our fiducial model with one in which Pop~III stars are added following our simple model with the same $f_{\rm esc}$ as more massive halos (magenta curves, only shown for Model~II). The formation of Pop~III stars effectively enhances the integrated ionizing efficiency, so that the reionization history develops a long tail toward higher redshifts. In our simple model, it does so in such a way that the additional contribution is largely independent of the emissivity at the end of reionization, which provides one route toward simultaneously matching the ionizing emissivity at $z<6$ and $\tau_{\rm es}$. However, the optical depth measurements still require the efficiency of star formation in these halos to be relatively low.

\begin{table}
	\caption{
	Parameter choices for the sample cases shown in Fig.~\ref{fig:swf}, which include a step-wise evolution of $f_{\rmn{esc}}$.}
	\label{tb:swf_cases}
	\begin{tabular}{@{}cccccc}
	\hline
	\hline
	Case & $f_{\rmn{esc,1}}$ & $f_{\rmn{esc,2}}$ & $M_{\rmn{min}}$ $^{\rmn{i}}$ [$M_{\odot}$] & Sup & PopIII \\
	\hline
	1 & 0.5 & 0.01 & $3.6 \times 10^{8}$ & No & No \\	
	2 & 0.5 & 0.1 & $3.6 \times 10^{8}$ & Yes & $z_{\rmn{end}}=6$ \\
	3 & 0.3 & 0.05 & $3.6 \times 10^{8}$ & No & No \\
	4 & 0.3 & 0.1 & $1.1 \times 10^{8}$ & Yes & $z_{\rmn{end}}=6$ \\
	\hline
	\end{tabular}

 \medskip
$^{\rmn{i}}$ 
This is evaluated at $z=7$.
\end{table}

\begin{figure*}
 \includegraphics[width=0.75\textwidth]{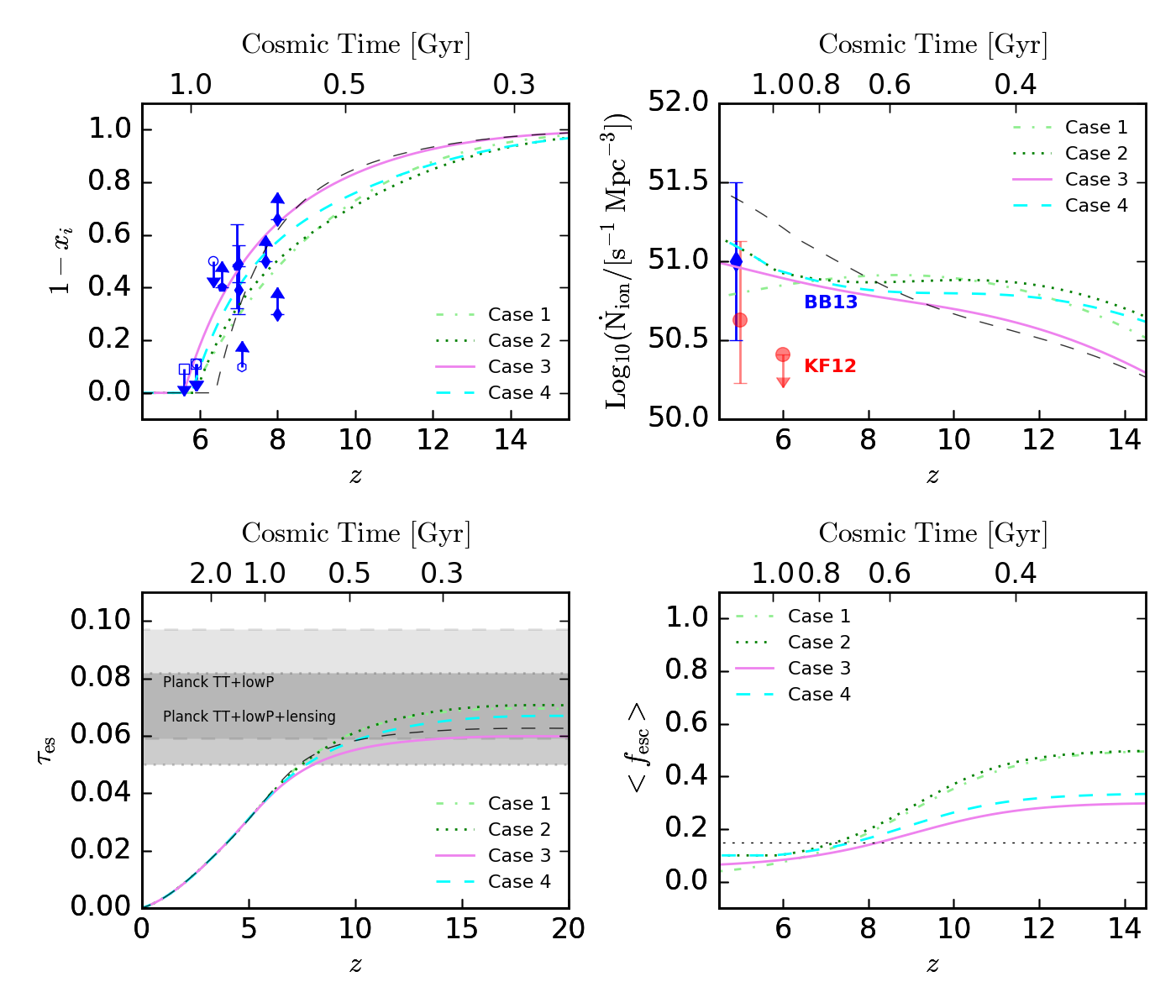}
 \caption{
The upper and lower left panels are identical to those in Fig.~\ref{fig:w_advanced_model}, except using the variations of Model~II listed in Table~\ref{tb:swf_cases}. 
The lower right panel shows the time evolution of the escape fraction, $\langle f_{\rmn{esc}} \rangle$, averaged over halo mass. The thin dashed curves are taken from Fig.~\ref{fig:w_advanced_model}, assuming both photoionization suppression and the formation of PopIII stars. Allowing $f_{\rmn{esc}}$ to evolve with halo mass (i.e. cosmic time) effectively reduces the post-reionization UV background. }
 \label{fig:swf}
\end{figure*}

\subsection{Evolution in $f_{\rmn{esc}}$}

As mentioned earlier, we find that the preferred parameter spaces for Models~II and III only marginally agree with the UV ionizing background measured at $z\sim5$. This happens because the two cases are dominated by star formation in relatively massive halos, whose abundance is increasing very rapidly (whether or not photo-heating suppression is included). The post-reionization ionizing background is always dominated by these massive halos and is sensitive to the escape fraction assigned to those massive halos. Thus making the escape fraction either a function of halo mass or of redshift can potentially reconcile the tension.

To this end we consider a simple method that was used to match to the higher $\tau_{\rmn{es}}$ measured by WMAP (see \citealt{Alvarez_2012}), which assigns different $f_{\rmn{esc}}$ to halos according their masses. Effectively, this allows $f_{\rmn{esc}}$ to evolve with time as well, because of evolution in the halo mass function. Fig.~\ref{fig:swf} shows a few sample cases based on Model~II (and specified in Table~\ref{tb:swf_cases}) that assume a stepwise evolution of $f_{\rmn{esc}}$ with halo mass. For the sake of simplicity, we assume the critical mass for $f_{\rmn{esc}}$ evolution to be the same as the critical mass scale, $M_{\rmn{c}}=5.0\times10^{9}M_{\odot}$, for photo-suppressible halos and assign two distinct escape fractions, $f_{\rmn{esc,1}}$ and $f_{\rmn{esc,2}}<f_{\rmn{esc,1}}$, to halos with mass below 
(the escape fraction of molecular halos is fixed at 0.5)
and above $M_{\rmn{c}}$, respectively. All of these choices reduce the emissivity at $z=5$ so as to be consistent with observations while remaining consistent with all or nearly all of our other constraints. They also increase the optical depth to electron scattering so that the models agree better with the best-fit result from Planck. 

The curves in the lower right panel of Fig.~\ref{fig:swf} show the time evolution the mean escape fraction, which is averaged by the total number of ionizing photons produced by star formation. By assumption, there will be no star formation in photo-sensitive halos after reionization ends. Consequently, the low $f_{\rmn{esc,2}} \la 0.1$ assigned to photo-insensitive halos yields a moderate emissivity at $z\sim5$, independent of the $f_{\rmn{esc,1}}$ of photo-sensitive halos. On the contrary, the much higher $f_{\rmn{esc,1}}$ quickly elevates the mean escape fraction to the level required to reionize the IGM by $z\sim6$ ($\sim0.15$ for Model~II as indicated by the horizontal dotted line, see also Fig.~\ref{fig:2D_color_plots}) as we trace back from the end of reionization, when photo-sensitive halos begin to contribute.


\section{Summary}

We have presented a model of the star formation efficiency of galaxies during reionization developed from the abundance matching technique and the assumption that galactic growth is driven by continuous mass accretion of dark matter halos. A variety of existing observations allow us to constrain the parameters of our models of cosmic star formation and reionization, shedding light onto the ionizing efficiency of known galaxy populations and their influence on reionization. 

In contrast to many previous works, which simply treat the star formation efficiency as a constant, we are able to more accurately model its evolution with halo mass and redshift by applying the halo abundance matching technique to the dust-corrected, UV luminosity functions at $z=5$--8. We find that $f_{\star}$ evolves strongly with halo mass at $z>5$, in qualitative agreement with its evolution at lower redshifts, and peaks at a characteristic halo mass 
$\log_{10}(M_{\rmn{peak}}/M_{\odot})=-0.15(1+z)+12.8$. We model the evolution of $\log f_{\star}$ in halos more massive than $2 \times 10^{10} M_{\odot}$ with a third order polynomial in $\log M$ as given by equation~(\ref{eq:fit_f_mz}) and investigate different extrapolations to lower masses. The mass to light ratio derived from our models is in good agreement with observations. Because the luminosity function in this period is so steep, the behavior of low-mass halos is particularly important. Recent data from the Hubble Frontier Field \citep{Atek_2015} suggests a rapid decline in the star formation efficiency below the peak, with $f_\star \propto M^{1/2}$, albeit with large uncertainties.  We therefore allow for a range of extrapolations to small halo masses.

By empirically extrapolating our model of $f_{\star}(M,z)$ to both smaller masses and earlier times, we predict the UV luminosity function and the star formation rate density at $z=9$--10 in reasonable agreement with current observations. Our fiducial power-law extrapolations of $f_{\star}$ in low-mass halos suggests a roughly linear or even slightly flattening $\rho_{\rmn{SFR}}$ in the log--log space between $6<z<12$, with little evidence of rapid evolution at $z \ga 8$. We note, however, that the evolution pattern can be complicated by the dust correction, the assumed extrapolation of $f_{\star}$, and the underlying model of galactic growth. We also provide forecasts for future JWST observations based on our star formation histories. With a $10^{6}$~s observation, JWST is expected to measure the faint-end slope of UV luminosity function to $M_{\rmn{lim}}\sim-16$. For a power-law extrapolation of $\log f_{\star}$ with slope 0.5, we predict that JWST may detect 10 UV-bright galaxy per 10 $\rmn{arcmin^{2}}$ beyond $z\sim15$ and more than half of the total UV luminosity emitted at $z\sim8$. 

The qualitative behavior of $f_{\star}(M,z)$, with a peak at 
$\sim 30\%$ near $M_h \sim 10^{11}$--$10^{12}$~M$_\odot$, is similar to analogous studies at lower redshifts. There, the decline toward higher masses is attributed to a combination of gravitational shock heating and feedback from active galactic nuclei (AGNs), but these effects have not been extensively studied at high redshifts. Any explanation of this trend at high masses will have to account for both the clear (though gentle) redshift dependence of the peak, which moves to smaller masses at higher redshifts, and the \emph{decreasing} star formation efficiency at fixed mass as redshift increases.

At small halo masses, the star-formation efficiency is usually assumed to be set by stellar feedback, either through supernovae or winds and radiation pressure. Those processes certainly operate at these high redshifts, and the shape inferred from the \citet{Atek_2015} data is consistent with expectations from simple models. This picture is also consistent with the increasing star formation efficiency in faint halos toward higher redshifts, as (at a given mass) the binding energy of halos increases with redshift.

We emphasize that our model is intentionally simple, ignoring a number of complications that must be addressed in the future. We associate star formation with halo accretion and ignore stochasticity in that relation (including mergers and a duty cycle for star formation). We also assume that the stellar populations remain constant within halos (including the IMF and metallicity).  Evolution in any of these parameters could explain some or all of the trends we see: for example, if massive galaxies have significantly higher metallicities or less top-heavy IMFs than small galaxies, our model would erroneously assign them a smaller star formation efficiency. But the trend would have to be quite strong to account for all of the effect we see. 

Modeling $f_{\star}$ as a function of both halo mass and redshift, we find a slower evolution of the cosmic star formation rate density compared to that predicted by models that fix the star formation efficiency as a function of redshift (see also \citealt{Mashian_2016}; \citealt{Mason_2015}). Consequently, it is easier to explain the observed electron scattering optical depth using the star formation histories predicted by our redshift-dependent models, especially if one desires a small escape fraction comparable to that observed at lower redshifts. Our baseline models show that with a single average $f_{\rmn{esc}}$ of $\la 0.2$ and a minimum halo mass of $\sim 10^{8} M_{\odot}$, known galaxy populations are able to ionize the IGM by $z \sim 6$ and reproduce the optical depth recently observed by the Planck satellite. However, the ionizing background at $z=5$ might be overestimated, especially when star formation in massive halos dominates the emissivity (Models~II and III). We consequently explore the possibilities of photoionization heating, massive Pop~III stars, and evolution in $f_{\rm esc}$, in addition to our baseline model. We find that photoheating can only be a significant feedback mechanism when the shape of $f_{\star}(M) \times f_{\rm esc}$ differs \emph{qualitatively} from that at low redshifts, allowing faint galaxies to contribute a substantial fraction of ionizing photons. Our crude analysis of Pop~III stars demonstrates that their presence effectively enhances the ionizing efficiency while having little impact on the post-reionization emissivity, which causes a tail of ionization to high redshifts and improves the agreement with the electron scattering optical depth. Meanwhile, the recent $\tau_{\rmn{es}}$ measurement from Planck places an upper limit $f_{\rmn{*,m}} \la 10^{-3}$ on the efficiency at which those earliest stars could form (see also \citealt{Visbal_2015}). Finally, the tension between our predicted emissivity of ionizing photons and that estimated from observations might be resolved by allowing the escape fraction to evolve with halo mass or redshift (see also e.g., \citealt{Alvarez_2012}). Assuming that the escape fraction is typically higher in faint galaxies, we suggest that bright galaxies with an absolute escape fraction less than 10\% can reproduce the observed reionization history without over-predicting the emissivity or relying on efficient star formation in low-mass halos.  

Finally, we must stress that the models of $f_{\star}$ presented in this paper demonstrate that the known galaxy population can self-consistently reproduce the histories of both galactic growth and cosmic reionization using a simple prescription for the halo assembly history. In the future, improved observations and numerical simulations with careful treatments of feedback will improve our understanding of the star formation efficiency trends with halo mass and redshift. 

\section*{Acknowledgments}
The authors would like to thank 
the anonymous referee, as well as
Richard Ellis, Phil Hopkins, Yu Lu, Jordan Mirocha, and Brant Robertson 
for extremely helpful comments on the manuscript. GS also thanks Jamie Bock for his kind support and Jianfei Shen for discussions of the statistical inference. This research was completed as part of the University of California Cosmic Dawn Initiative. We acknowledge support from the University of California Office of the President Multicampus Research Programs and Initiatives through award MR-15-328388.  SRF was partially supported by NASA grant NNX15AK80G, administered through the ATP program, and by a Simons Fellowship in Theoretical Physics. SRF also thanks the Observatories of the Carnegie Institute of Washington for hospitality while much of this work was completed.

\bibliographystyle{./mn2e}

\bibliography{./constr}

\bsp

\label{lastpage}

\end{document}